\documentclass[aps, prc, floatfix, nofootinbib, superscriptaddress, twocolumn]{revtex4-1}

\usepackage{latexsym}
\usepackage{amsmath}
\usepackage{amssymb}
\usepackage{amsfonts}
\usepackage{makecell}

\usepackage[mathscr,scaled=1.15]{urwchancal}
\DeclareFontFamily{OT1}{pzc}{}
\DeclareFontShape{OT1}{pzc}{m}{it}%
{<-> s * [1.15] pzcmi7t}{}
\DeclareMathAlphabet{\mathpzc}{OT1}{pzc}{m}{it}

\usepackage{color}

\usepackage{supertabular}
\usepackage{placeins}
\usepackage{epsfig}
\usepackage{graphicx}
\usepackage{multirow}

\definecolor{purple}{rgb}{0.5,0,0.5}
\definecolor{blue}{rgb}{0.0,0,0.9}
\definecolor{prdblue}{rgb}{0.133,0.118,0.498}
\usepackage[colorlinks=true, pdfstartview=FitV, linkcolor=prdblue, citecolor= prdblue, urlcolor=prdblue]{hyperref}

\begin{document}

\title{Doubly heavy tetraquark states $cc\bar{u}\bar{d}$ and $bb\bar{u}\bar{d}$}

\author{Xiaoyun Chen}
\email[]{xychen@jit.edu.cn}
\affiliation{College of Science, Jinling Institute of Technology, Nanjing 211169, P. R. China}

\author{Youchang  Yang}
\email[]{yangyc@gues.edu.cn}
\affiliation{Guizhou University of Engineering Science; Zunyi Normal University, P. R. China}

\begin{abstract}
Inspired by the recent observation of a very narrow state, called $T_{cc}^+$, by the LHCb collaboration, the possible bound states and low-lying
resonance states of the doubly heavy tetraquark states $cc\bar{u}\bar{d}$ ($T_{cc}$) and $bb\bar{u}\bar{d}$ ($T_{bb}$) are searched in the
framework of chiral quark model with an accurate few body method, Gaussian expansion method \cite{GEM}. The real scaling method \cite{rs2} is also applied to identify
the genuine resonance states. In the calculation, the meson-meson structure, diquark-antidiquark structure, as well as their coupling are all
considered. The numerical results show that (i) for $T_{cc}$ and $T_{bb}$, only $I(J^P)=0(1^+)$ states are bound in different quark structures.
The binding energy varies from a few MeV for meson-meson structure to over 100 MeV for diquark-antidiquark structure. For example for $T_{cc}$,
in meson-meson structure, there exists a weakly bound molecule $DD^*$ state around 3841.4 MeV, 1.8 MeV below the $D^0D^{*+}$, which may be a good
candidate of the observed state by LHCb; however in diquark-antidiquark structure, a deeper bound sate with mass 3700.9 MeV is obtained;
when considering the structure mixing, the energy of system decreases to 3660.7 MeV and the shallow bound state disappears.
(ii) Besides bound states, several resonance states for $T_{QQ}(Q=c, b)$ with $I(J^P)=1(0^+), 1(1^+), 1(2^+), 0(1^+)$ are proposed.
\end{abstract}

\maketitle


\section{Introduction}
\label{introduction}
Recently, the LHCb collaboration \cite{LHCb:2021vvq} reported the observation of a very narrow state, called $T_{cc}^+$, in the $D^0D^0\pi^+$ invariant mass
spectrum. The binding energy and the decay width is:
\begin{eqnarray}
\delta m_{\rm{BW}}&=&-273\pm61\pm5^{+11}_{-14}\quad \rm{kev/c^2}, \nonumber\\
\Gamma_{\rm{BW}}&=&410\pm165\pm43^{+18}_{-38} \quad\rm{kev}.
\end{eqnarray}
The LHCb Collaboration also released a decay analysis, in which the unitarised Breit-Wigner profile was used  \cite{LHCb:2021auc}. The mass with respect to the $D^{*+}D^0$ threshold and width reads,
\begin{eqnarray}
\delta m^{U}&=&-360\pm40^{+4}_{-0}\quad \rm{kev/c^2}, \nonumber\\
\Gamma^{U}&=&48\pm2^{+0}_{-14} \quad\rm{kev}.
\end{eqnarray}
This observation has two points worth our attention. Firstly, different from the hidden charm or hidden bottom exotic hadron states observed
experimentally before, this is the first observation of an exotic state with open double charm. Date back to the year of 2002, SELEX collaboration first
reported the observation of the doubly charmed baryon $\Xi_{cc}^+ (ccd)$ in the channels $\Lambda_c^+K^-\pi^+$ and $pD^+K^-$ \cite{SELEX:2002wqn,SELEX:2004lln}.
Fifteen years later, LHCb collaboration also found the doubly charmed baryon $\Xi_{cc}^{++}(ccu)$ in the $\Lambda_c^+K^-\pi^+\pi^+$ mass distribution
\cite{LHCb:2017iph}. The value of the mass is 100 MeV higher than $\Xi_{cc}^+$ observed by SELEX collaboration. Secondly, the mass of observed $T_{cc}$ is just
a litter lower than $M(D^0)+M(D^{*+})$, with a very small binding energy. Undoubtedly, the observation of $T_{cc}^+$ will open a brand new window to search
for the new hadron states beyond the traditional hadrons, both experimentally and theoretically.

The discovery of $T_{cc}^+$ and $\Xi_{cc}^{++}$ has important impact since it indicates that two identical charm quarks can exist in a hadronic state,
and inspires some theoretical studies on the possible doubly charmed tetraquarks and its partner states doubly bottomed tetraquarks. Historically,
the first study of $QQ\bar{q}\bar{q}$ was made in early 1980s \cite{Ader:1981db}, with the observation that the system will be bound, below the
$Q\bar{q}+Q\bar{q}$ threshold if the mass ratio $M/m$ becomes large enough. This was confirmed by Heller \emph{et al.} \cite{Heller:1986bt,Carlson:1987hh}
and Zouzou \emph{et al.} \cite{Zouzou:1986qh}. The first phenomenological attempt to estimate doubly heavy tetraquark mass was carried out by Lipkin
using the nonrelativistic quark model in 1986 \cite{lipkin}. The author pointed out $M_{T_{cc}} \leq 3935$ MeV, 60 MeV above the threshold, and $T_{bb}$
was a bound state with the binding energy of 224 MeV.

Until now, there are so many literatures about the doubly heavy tetraquarks $QQ\bar{q}\bar{q}  (Q=c, b ; q=u, d, s)$. Such as, the color-magnetic interaction
(CMI) model \cite{Lee:2009rt,Hyodo:2012pm,Luo:2017eub,Li:2012ss,Guo:2021yws,Weng:2021hje},
quark models \cite{EPJA19,Karliner:2017qjm,wpark,Hernandez:2019eox,Yang:2019itm,Yang:2009zzp,Tan:2020ldi,Lu:2020rog,Ebert:2007rn,Ren:2021dsi,Xing:2018bqt,Yan:2018gik,Richard:2018yrm,Carames:2011zz,Meng:2020knc,Wang:2021yld,Zhang:2021yul,Noh:2021lqs,Deng:2018kly},
QCD sum rule approach \cite{Du:2012wp,Wang:2017dtg,Tang:2019nwv,Navarra:2007yw,Agaev:2021vur,Agaev:2018khe,Agaev:2020mqq,Agaev:2019qqn,Agaev:2020dba,Azizi:2021aib,Xin:2021wcr} lattice QCD simulation \cite{Brown:2012tm,Ikeda:2013vwa,Bicudo:2012qt,Bicudo:2015vta,Francis:2016hui,Junnarkar:2018twb,Leskovec:2019ioa,Hudspith:2020tdf,Mohanta:2020eed}, effective field theory \cite{Meng:2021jnw,Ling:2021bir,Dong:2021bvy} and so on \cite{Qin:2020zlg,Ali:2018ifm,Ali:2018xfq}.
One of the controversies is whether $QQ\bar{q}\bar{q}$ tetraquarks with two heavy quarks $Q$ and two light antidiquarks $\bar{q}$ are stable or not
against the decay into two $Q\bar{q}$ mesons. Actually, this dispute has been a long history due to the lack of experiment information about the strength
of the interaction between two heavy quarks. The other important question is if the $T_{QQ}$ is bound, is it tightly bound or loosely bound?

Most of theoretical calculations predict the double bottom tetraquark states, at least $1^+$ states, lie below the open bottom threshold. Whereas for doubly
charmed tetraquarks, some work support they are above the open charm threshold \cite{Karliner:2017qjm,Ebert:2007rn,Du:2012wp,Wang:2017dtg,Navarra:2007yw,Ikeda:2013vwa,Cheng:2020wxa}. In Ref. \cite{Cheng:2020wxa}, the authors stated
$T_{cc}$ was a $I(J^P)=0(1^+)$ state around 3929 MeV (53 MeV above the $DD^*$ threshold) and all the double-charm tetraquarks were not stable.
Karliner \emph{et al.}~\cite{Karliner:2017qjm} predicted the mass of $T(cc\bar{u}\bar{d})$ with $J^P=1^+$ to be 3882 MeV, 7 MeV above the $D^0D^{*+}$
threshold and 148 MeV above $D^0D^+\gamma$ threshold against the strong and weak decays. In lattice QCD simulation, the authors \cite{Ikeda:2013vwa}
manifested that the phase shifts in the isospin triplet ($I=1$) channels indicated repulsive interactions, while those in the $I=0$ channels suggested
attraction, although neither bound states nor resonance states were found in the $T_{cc} (IJ^P=01^+)$. Some work are in favor of them as tightly bound states \cite{Lee:2009rt,Hyodo:2012pm,Luo:2017eub,Tan:2020ldi,Deng:2018kly}. The common feature of researches obtaining a deeply bound state is that the diquark-antidiquark structure is
employed. For example, in Ref. \cite{Hyodo:2012pm}, the mass splitting indicated that the mass of $T_{cc}$ with color structure $6\otimes\bar{6}$ lied
above $DD^*$ threshold, but the mass of $T_{cc}$ with color structure $\bar{3}\otimes3$ lied at 71 MeV below $DD^*$ threshold.
Whereas in Ref.~\cite{Li:2012ss}, Li \emph{et al.} got a loosely bound molecule state with 470 keV binding energy, which was consistent with the recent
experimental data~\cite{LHCb:2021vvq,LHCb:2021auc}. In their work, only meson-meson structure was considered.
These theoretical work infers that color structures and quark-quark interactions may play an important role in the $T_{QQ}$ states.

So discovery of $T_{cc}$ state provides a chance to check the quark-quark interactions for various theoretical approaches based on quark degree of
freedom. In quark model, the quark-quark interactions within confinement scale ($\sim$ 1 fm) have undergone a wide check in the hadron spectrum, where the unique color structure, singlet, is accepted.
When we apply quark-quark interactions to multiquark systems, the ``Casimir scaling" is employed for generalization~\cite{Casimir}, although this generalization may cause anti-confinement in the multiquark
systems~\cite{anticonfine}. In ``Casimir scaling" scheme, the two-body interactions used in color singlet $qqq$ and $q\bar{q}$ systems are directly
extended to quark-pair with various color structures, and the effects of color structure are taken care by Casimir operator
$\boldsymbol{\lambda}_i\cdot \boldsymbol{\lambda}_j$.

With the accumulation of experimental data on multiquark system, it is time to check ``Casimir scaling" in detail. Herein, we apply the chiral quark
model (ChQM) to the tetraquark system $T_{QQ}$ with meson-meson and diquark-antidiquark structures, and  generalize the quark-quark
interactions used in color singlet baryons and mesons to multiquark system by ``Casimir scaling". The contributions of each term in Hamiltonian
for different color structures are extracted, and shown to study the effects of color structure. In this way, we are trying to make clear why
the diquark-antiquark structure leads to deeply bound states, whereas meson-meson structure brings about weakly bound states. In the present work, the doubly heavy tetraquarks $cc\bar{u}\bar{d}$ and $bb\bar{u}\bar{d}$ with the quantum numbers $I(J^P)=1(0^+), 1(1^+), 1(2^+), 0(1^+)$ constraint of the Pauli principle
in the framework of the ChQM are investigated. Single channel and various channel coupling calculations are performed to show the influence of color
structure. Meanwhile, the possible resonance states are also searched within a real scaling method in the complete coupled-channels.

The paper is organized as follows. In the next section, we present the chiral quark model (ChQM) and the accurate few-body computing way: Gaussian expansion
method(GEM), as well as the wave functions of the four-body $T_{QQ}$ system. In Sec. \ref{results}, we present and analyze our results.
Finally, a summary is given in the last section.



\section{Theoretical framework}
\label{ModleandGEM}
\subsection{The chiral quark model}
Many theoretical methods have been used to uncover the properties of multiquark candidates observed by experiments since 2003. One of them, the QCD-inspired quark model, is still the effective and simple tool to describe the hadron spectra and hadron-hadron interactions and has been witnessed great achievements. It has been used in our previous work to investigate the tetraquark systems and got some helpful information \cite{Chen:2016npt,Chen:2018hts,Chen:2019vrj}. Herein, the application of the chiral quark model in doubly heavy tetraquark states $T_{QQ}$ is quite expected.

The Hamiltonian of the chiral quark model can be written as follows for four-body system,
\begin{align}
 H & = \sum_{i=1}^4 m_i  +\frac{p_{12}^2}{2\mu_{12}}+\frac{p_{34}^2}{2\mu_{34}}
  +\frac{p_{1234}^2}{2\mu_{1234}}  \quad  \nonumber \\
  & + \sum_{i<j=1}^4 \left[ V_{ij}^{C}+V_{ij}^{G}+\sum_{\chi=\pi,K,\eta} V_{ij}^{\chi}
   +V_{ij}^{\sigma}\right].
\end{align}
The potential energy: $V_{ij}^{C, G, \chi, \sigma}$ represents the confinement, one-gluon-exchange(OGE), Goldston boson exchange and scalar $\sigma$ meson-exchange,
respectively. According to Casimir scheme, the forms of these potentials can be directly extended to multiqaurk systems with the Casimir factor
$\boldsymbol{\lambda}_i \cdot \boldsymbol{\lambda}_j $~\cite{Casimir}. The forms of them are:
{\allowdisplaybreaks
\begin{subequations}
\begin{align}
V_{ij}^{C}&= ( -a_c r_{ij}^2-\Delta ) \boldsymbol{\lambda}_i^c
\cdot \boldsymbol{\lambda}_j^c ,  \\
 V_{ij}^{G}&= \frac{\alpha_s}{4} \boldsymbol{\lambda}_i^c \cdot \boldsymbol{\lambda}_{j}^c
\left[\frac{1}{r_{ij}}-\frac{2\pi}{3m_im_j}\boldsymbol{\sigma}_i\cdot
\boldsymbol{\sigma}_j
  \delta(\boldsymbol{r}_{ij})\right],  \\
\delta{(\boldsymbol{r}_{ij})} & =  \frac{e^{-r_{ij}/r_0(\mu_{ij})}}{4\pi r_{ij}r_0^2(\mu_{ij})}, \\
V_{ij}^{\pi}&= \frac{g_{ch}^2}{4\pi}\frac{m_{\pi}^2}{12m_im_j}
  \frac{\Lambda_{\pi}^2}{\Lambda_{\pi}^2-m_{\pi}^2}m_\pi v_{ij}^{\pi}
  \sum_{a=1}^3 \lambda_i^a \lambda_j^a,  \\
V_{ij}^{K}&= \frac{g_{ch}^2}{4\pi}\frac{m_{K}^2}{12m_im_j}
  \frac{\Lambda_K^2}{\Lambda_K^2-m_{K}^2}m_K v_{ij}^{K}
  \sum_{a=4}^7 \lambda_i^a \lambda_j^a,   \\
\nonumber
V_{ij}^{\eta} & =
\frac{g_{ch}^2}{4\pi}\frac{m_{\eta}^2}{12m_im_j}
\frac{\Lambda_{\eta}^2}{\Lambda_{\eta}^2-m_{\eta}^2}m_{\eta}
v_{ij}^{\eta}  \\
 & \quad \times \left[\lambda_i^8 \lambda_j^8 \cos\theta_P
 - \lambda_i^0 \lambda_j^0 \sin \theta_P \right],   \\
 v_{ij}^{\chi} & =  \left[ Y(m_\chi r_{ij})-
\frac{\Lambda_{\chi}^3}{m_{\chi}^3}Y(\Lambda_{\chi} r_{ij})
\right]
\boldsymbol{\sigma}_i \cdot\boldsymbol{\sigma}_j,\\
V_{ij}^{\sigma}&= -\frac{g_{ch}^2}{4\pi}
\frac{\Lambda_{\sigma}^2}{\Lambda_{\sigma}^2-m_{\sigma}^2}m_\sigma \nonumber \\
& \quad \times \left[
 Y(m_\sigma r_{ij})-\frac{\Lambda_{\sigma}}{m_\sigma}Y(\Lambda_{\sigma} r_{ij})\right]  ,
\end{align}
\end{subequations}}
\hspace*{-0.5\parindent}%
where $Y(x)  =   e^{-x}/x$;
$\{m_i\}$ are the constituent masses of quarks and antiquarks, and $\mu_{ij}$ are their reduced masses;
\begin{equation}
\mu_{1234}=\frac{(m_1+m_2)(m_3+m_4)}{m_1+m_2+m_3+m_4};
\end{equation}
$\mathbf{p}_{ij}=(\mathbf{p}_i-\mathbf{p}_j)/2$, $\mathbf{p}_{1234}= (\mathbf{p}_{12}-\mathbf{p}_{34})/2$;
$r_0(\mu_{ij}) =s_0/\mu_{ij}$;
$\boldsymbol{\sigma}$ are the $SU(2)$ Pauli matrices;
$\boldsymbol{\lambda}$, $\boldsymbol{\lambda}^c$ are $SU(3)$ flavor, color Gell-Mann matrices, respectively;
$g^2_{ch}/4\pi$ is the chiral coupling constant, determined from the $\pi$-nucleon coupling;
and $\alpha_s$ is an effective scale-dependent running coupling \cite{Valcarce:2005em},
\begin{equation}
\alpha_s(\mu_{ij})=\frac{\alpha_0}{\ln\left[(\mu_{ij}^2+\mu_0^2)/\Lambda_0^2\right]}.
\end{equation}
All the parameters are determined by fitting the meson spectra, from light to heavy; and the resulting values are listed in Table~\ref{modelparameters}.

\linespread{1.2}
\begin{table}[!t]
\begin{center}
\caption{ \label{modelparameters}
Model parameters, determined by fitting the meson spectra.}
\begin{tabular}{llr}
\hline\hline\noalign{\smallskip}
Quark masses   &$m_u=m_d$    &313  \\
   (MeV)       &$m_s$         &536  \\
               &$m_c$         &1728 \\
               &$m_b$         &5112 \\
\hline
Goldstone bosons   &$m_{\pi}$     &0.70  \\
   (fm$^{-1} \sim 200\,$MeV )     &$m_{\sigma}$     &3.42  \\
                   &$m_{\eta}$     &2.77  \\
                   &$m_{K}$     &2.51  \\
                   &$\Lambda_{\pi}=\Lambda_{\sigma}$     &4.2  \\
                   &$\Lambda_{\eta}=\Lambda_{K}$     &5.2  \\
                   \cline{2-3}
                   &$g_{ch}^2/(4\pi)$                &0.54  \\
                   &$\theta_p(^\circ)$                &-15 \\
\hline
Confinement        &$a_c$ (MeV fm$^{-2}$)         &101 \\
                   &$\Delta$ (MeV)     &-78.3 \\
\hline
OGE                 & $\alpha_0$        &3.67 \\
                   &$\Lambda_0({\rm fm}^{-1})$ &0.033 \\
                  &$\mu_0$(MeV)    &36.98 \\
                   &$s_0$(MeV)    &28.17 \\
\hline\hline
\end{tabular}
\end{center}
\end{table}

With these model parameters, we get the relevant meson spectra $D^{(*)}$ and $B^{(*)}$ list in Table ~\ref{mesonspectra}. In comparison with experiments, we can see that the quark model can successfully describe the hadron spectra.

\linespread{1.2}
\begin{table}[!t]
\begin{center}
\renewcommand\tabcolsep{6.0pt} 
\caption{ \label{mesonspectra} The mass spectra of $c\bar{u}$, $c\bar{d}$, $b\bar{u}$, $b\bar{d}$ in the chiral quark model in comparison with the experimental data \cite{PDG} (in unit of MeV).}
\begin{tabular}{ccccc}
\hline\hline\noalign{\smallskip}
State      &Meson &$I(J^P)$  &Energy &Expt \cite{PDG}\\
\hline
$c\bar{u}$ &$D^0$ &$\frac{1}{2}(0^-)$  &1862.6     &1864.8 \\
                         &$D^{*0}$ &$\frac{1}{2}(1^-)$  &1980.6        &2007.0\\
$c\bar{d}$ &$D^+$ &$\frac{1}{2}(0^-)$  &1862.6     &1869.6 \\
                         &$D^{*+}$ &$\frac{1}{2}(1^-)$  &1980.6        &2010.3\\
$b\bar{u}$ &$B^-$ &$\frac{1}{2}(0^-)$  &5280.8     &5279.3 \\
                         &$B^{*-}$ &$\frac{1}{2}(1^-)$  &5319.6        &5325.2\\
$b\bar{d}$ &$\bar{B}^0$ &$\frac{1}{2}(0^-)$  &5280.8     &5279.6 \\
                         &$\bar{B^*}^0$ &$\frac{1}{2}(1^-)$  &5319.6        &5325.2\\
\hline\hline
\end{tabular}
\end{center}
\end{table}

\subsection{Wave functions of $T_{QQ}$}
\begin{figure}
\resizebox{0.40\textwidth}{!}{\includegraphics{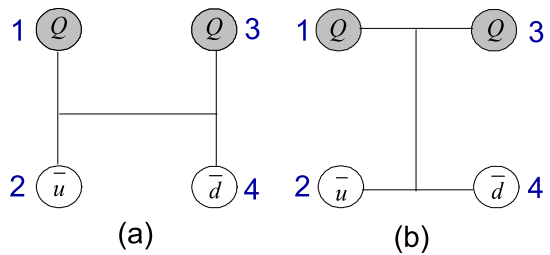}}
\caption{\label{structure} Two types of configurations of $T_{QQ}~(Q = c, b)$. Figure (a) represents the meson-meson structure; figure (b) is the diquark-antidiquark structure. }
\end{figure}

For $T_{QQ}~(Q = c, b)$ system, there are two quark configurations, meson-meson structure (MM) and diquark-antidiquark structure (DA), which are shown in Fig.\ref{structure}. Both of them and their coupling effects are considered in this work.

The wave functions of tetraquark states should be a product of spin, flavor, color and space degrees of freedom. For spin part, We denote $\alpha$ and $\beta$ as the spin-up and spin-down states of quarks and the spin wave functions for two-quark system take,
\begin{align}
&\chi_{11}=\alpha\alpha,~~
\chi_{10}=\frac{1}{\sqrt{2}}(\alpha\beta+\beta\alpha),~~
\chi_{1-1}=\beta\beta,\nonumber \\
&\chi_{00}=\frac{1}{\sqrt{2}}(\alpha\beta-\beta\alpha),
\end{align}
then total six wave functions of four-body system are obtained easily, which are shown in Table \ref{wavefunctions}. The subscript of $\chi$ represents the spin values $S_1$ and $S_2$ of two sub-clusters. The superscript of $\chi$ stands for the total spin $S$ ($S_1 \otimes S_2$ = S) and the third projection $M_S$ of $S$ for four-quark system.

For flavor part, the flavor wave functions with isospin $= 0$ and $1$, are also tabulated in Table \ref{wavefunctions}. The subscript $00$ or $11$ of $\chi$ represents the total isospin $I$ and the third projection $I_z$ of $I$.

For color part, more richer structures in four-quark system will be considered than conventional $q\bar{q}$ mesons and $qqq$ baryons.
For meson-meson structure in Fig. \ref{structure}, in the $SU(3)$ group, the colorless wave functions can be obtained from
$\mathbf{1}\otimes \mathbf{1}=\mathbf{1}$ or $\mathbf{8} \otimes \mathbf{8}=\mathbf{1}$. For diquark-antidiquark structure, the colorless wave
functions can be get from $\bar{\mathbf{3}} \otimes \mathbf{3}=\mathbf{1}$ or $\mathbf{6} \otimes \bar{\mathbf{6}}=\mathbf{1}$.
The detailed expressions of functions can be found in Table \ref{wavefunctions}.

\begin{table*}[!t]
\linespread{1.2}
\begin{center}
\caption{ \label{wavefunctions}The wave functions of spin, flavor, color part for $T_{QQ}$.}
\begin{tabular}{lll}
\hline\hline\noalign{\smallskip}
\quad \quad \quad Spin                                              & Flavor                   &\quad\quad \quad \quad \quad \quad \quad Color \\
\hline
$\chi_{00}^{00}=\chi_{00}\chi_{00}$
&$\chi^{00}=-\frac{1}{2}(c\bar{d}c\bar{u}-c\bar{u}c\bar{d})$
&\quad \quad $1 \otimes 1: \chi^{c1}=\frac{1}{3}(\bar{r}r+\bar{g}g+\bar{b}b)(\bar{r}r+\bar{g}g+\bar{b}b)$    \\

$\chi_{11}^{00}=\sqrt{\frac{1}{3}}(\chi_{11}\chi_{1-1}-\chi_{10}\chi_{10}+\chi_{1-1}\chi_{11})$
&$\chi^{11}=-\frac{1}{2}(c\bar{d}c\bar{u}+c\bar{u}c\bar{d})$
&\quad \quad \makecell[l]{$8 \otimes 8: \chi^{c2}=\frac{\sqrt{2}}{12}(3\bar{b}r\bar{r}b+3\bar{g}r\bar{r}g+3\bar{b}g\bar{g}b+3\bar{g}b\bar{b}g+3\bar{r}g\bar{g}r$\\
$\quad \quad \quad +3\bar{r}b\bar{b}r+2\bar{r}r\bar{r}r+2\bar{g}g\bar{g}g+2\bar{b}b\bar{b}b-\bar{r}r\bar{g}g$ \\
$\quad \quad \quad -\bar{g}g\bar{r}r-\bar{b}b\bar{g}g-\bar{b}b\bar{r}r-\bar{g}g\bar{b}b-\bar{r}r\bar{b}b)$}\\

$\chi_{01}^{11}=\chi_{00}\chi_{11}$
&
&\quad \quad \makecell[l]{$\bar{3} \otimes 3: \chi^{c3}=\frac{\sqrt{3}}{6}(rg\bar{r}\bar{g}-rg\bar{g}\bar{r}+gr\bar{g}\bar{r}-gr\bar{r}\bar{g}$ \\
$\quad \quad \quad+rb\bar{r}\bar{b}-rb\bar{b}\bar{r}+br\bar{b}\bar{r}-br\bar{r}\bar{b}$ \\
$\quad \quad \quad+gb\bar{g}\bar{b}-gb\bar{b}\bar{g}+bg\bar{b}\bar{g}-bg\bar{g}\bar{b})$}  \\

$\chi_{10}^{11}=\chi_{11}\chi_{00}$
&
&\quad \quad \makecell[l]{$6 \otimes \bar{6}:
\chi^{c4}=\frac{\sqrt{6}}{12}(2rr\bar{r}\bar{r}+2gg\bar{g}\bar{g}+2bb\bar{b}\bar{b}+rg\bar{r}\bar{g}+rg\bar{g}\bar{r}$\\
$\quad \quad \quad+gr\bar{g}\bar{r}+gr\bar{r}\bar{g}+rb\bar{r}\bar{b}+rb\bar{b}\bar{r}+br\bar{b}\bar{r}$  \\
$\quad \quad \quad+br\bar{r}\bar{b}+gb\bar{g}\bar{b}+gb\bar{b}\bar{g}+bg\bar{b}\bar{g}+bg\bar{g}\bar{b})$}\\

$\chi_{11}^{11}=\frac{1}{\sqrt{2}}(\chi_{11}\chi_{10}-\chi_{10}\chi_{11})$
&
&\makecell[l]{\quad\\\quad} \\

$\chi_{11}^{22}=\chi_{11}\chi_{11}$
&
& \\
\hline\hline
\end{tabular}
\end{center}
\end{table*}

Because of the quark contents of the present investigated four-quark systems are two identical heavy quarks ($Q=c, b$) and two identical light
antiquarks ($\bar{u}, \bar{d}$), the wave functions of $T_{QQ}$ should satisfy antisymmetry requirement. Then we collect all the color-spin bases
of $T_{QQ}$ states for possible quantum numbers according to the constraint from the Pauli principle in Table ~\ref{bases}.

\begin{table}[!t]
\linespread{1.2}
\begin{center}
\caption{ \label{bases}Color-spin bases for $T_{cc}$ system. The bases can be read as the notations: $[(c\bar{u})^{S1}_{c1}(c\bar{d})^{S2}_{c2}]^S$ for meson-meson structure, $[(cc)^{S1}_{c1}(\bar{u}\bar{d})^{S2}_{c2}]^S$ for diquark-antidiquark structure. The superscripts $S_1$, $S_2$ represent the spin for two sub-clusters; $S~(S_1 \otimes S_2=S)$ is the total spin of four-quark states. The subscripts $c_1$ and $c_2$ stand for color. For $T_{bb}$ sates, one just needs replace $c$ quark with $b$ quark simply.}
\begin{tabular}{ccccc}
\hline\hline\noalign{\smallskip}
$I(J^P)$           &$1(0^+)$ &$1(1^+)$ &$1(2^+)$ &$0(1^+)$ \\
\hline
\multirow{8}{*}{Bases}  &$[(c\bar{u})^0_1(c\bar{d})^0_1]^0$ &$[(c\bar{u})^0_1(c\bar{d})^1_1]^1$ &$[(c\bar{u})^1_1(c\bar{d})^1_1]^2$        &$[(c\bar{u})^0_1(c\bar{d})^1_1]^1$         \\
                        &$[(c\bar{u})^0_8(c\bar{d})^0_8]^0$ &$[(c\bar{u})^0_8(c\bar{d})^1_8]^1$ &$[(c\bar{u})^1_8(c\bar{d})^1_8]^2$        &$[(c\bar{u})^0_8(c\bar{d})^1_8]^1$             \\
                        &$[(c\bar{u})^1_1(c\bar{d})^1_1]^0$ &$[(c\bar{u})^1_1(c\bar{d})^0_1]^1$ &$[(cc)^1_{\bar{3}}(\bar{u}\bar{d})^1_3]^2$&$[(c\bar{u})^1_1(c\bar{d})^0_1]^1$                 \\
                        &$[(c\bar{u})^1_8(c\bar{d})^1_8]^0$ &$[(c\bar{u})^1_8(c\bar{d})^0_8]^1$ &                                          &$[(c\bar{u})^1_8(c\bar{d})^0_8]^1$                      \\
                        &$[(cc)^0_6(\bar{u}\bar{d})^0_{\bar{6}}]^0$ &$[(cc)^1_{\bar{3}}(\bar{u}\bar{d})^1_3]^1$       &                    &$[(c\bar{u})^1_1(c\bar{d})^1_1]^1$          \\
                        &$[(cc)^1_{\bar{3}}(\bar{u}\bar{d})^1_3]^0$ &&                                                                     &$[(c\bar{u})^1_8(c\bar{d})^1_8]^1$               \\
                        &                                           &&                                                                     &$[(cc)^0_6(\bar{u}\bar{d})^1_{\bar{6}}]^1$  \\
                        &                                           &&                                                                     &$[(cc)^1_{\bar{3}}(\bar{u}\bar{d})^0_3]^1$  \\
\hline\hline
\end{tabular}
\end{center}
\end{table}

Next, let's discuss the orbital wave functions for four-body system. They can be obtained by coupling the orbital wave function for each
relative motion of the system,
\begin{equation}\label{spatialwavefunctions}
\Psi_{L}^{M_{L}}=\left[[\Psi_{l_1}({\bf r}_{12})\Psi_{l_2}({\bf
r}_{34})]_{l_{12}}\Psi_{L_r}({\bf r}_{1234}) \right]_{L}^{M_{L}},
\end{equation}
where $l_1$ and $l_2$ is the angular momentum of two sub-clusters, respectively. $\Psi_{L_r}(\mathbf{r}_{1234})$ is the wave function of
the relative motion between two sub-clusters with orbital angular momentum $L_r$. $L$ is the total orbital angular momentum of four-quark states. In present work, we just consider the low-lying $S$-wave double heavy tetraquark states, so it is natural to assume that all the orbital angular
momenta are zeros.  In GEM, the spatial wave functions are expanded in series of Gaussian basis functions.
\begin{subequations}
\label{radialpart}
\begin{align}
\Psi_{l}^{m}(\mathbf{r}) & = \sum_{n=1}^{n_{\rm max}} c_{n}\psi^G_{nlm}(\mathbf{r}),\\
\psi^G_{nlm}(\mathbf{r}) & = N_{nl}r^{l}
e^{-\nu_{n}r^2}Y_{lm}(\hat{\mathbf{r}}),
\end{align}
\end{subequations}
where $N_{nl}$ are normalization constants,
\begin{align}
N_{nl}=\left[\frac{2^{l+2}(2\nu_{n})^{l+\frac{3}{2}}}{\sqrt{\pi}(2l+1)}
\right]^\frac{1}{2}.
\end{align}
$c_n$ are the variational parameters, which are determined dynamically. The Gaussian size
parameters are chosen according to the following geometric progression
\begin{equation}\label{gaussiansize}
\nu_{n}=\frac{1}{r^2_n}, \quad r_n=r_1a^{n-1}, \quad
a=\left(\frac{r_{n_{\rm max}}}{r_1}\right)^{\frac{1}{n_{\rm
max}-1}}.
\end{equation}
This procedure enables optimization of the expansion using just a small numbers of Gaussians. Finally, the complete channel wave function
$\Psi^{\,M_IM_J}_{IJ}$ for four-quark system is obtained by coupling the orbital and spin, flavor, color wave functions get in
Table \ref{bases}. At last, the eigenvalues of four-quark system are obtained by solving the Schr\"{o}dinger equation
\begin{equation}
    H \, \Psi^{\,M_IM_J}_{IJ}=E^{IJ} \Psi^{\,M_IM_J}_{IJ}.
\end{equation}

\section{Calculations and analysis}
\label{results}
In the present work, we are interested in looking for the bound states of low-lying $S$-wave $T_{QQ}~(Q=c, b)$ system. The allowed
quantum numbers are $I(J^P)=1(0^+), 1(1^+), 1(2^+), 0(1^+)$ with the constraint of Pauli principle. Meanwhile, the possible resonances of these states
are searched with the help of the real scaling method (RSM). For bound states calculations, we aim to study the influence of the color structures on the binding energy.

\begin{table}[!t]
\linespread{1.2}
\begin{center}
\caption{ \label{boundresults1} The low-lying eigenvalues of iso-vector $T_{QQ}$ states with $I(J^P)=1(0^+), 1(1^+), 1(2^+)$ (in unit of MeV). MM and DA represents the dimeson structures and diquark-antidiquark structures. $E_{cc}$ is the energy considering the coupling of MM and DA. $E_{Theo}$ and $E_{Exp}$ is the theoretical and experimental threshold. }
\begin{tabular}{ccccccc}
\hline\hline\noalign{\smallskip}
                              &$I(J^P)$   &MM     &DA     &$E_{cc}$  &$E_{Theo}$  &$E_{Exp}$   \\
                              \hline
\multirow{3}{*}{$T_{cc}$}       &$1(0^+)$   &3726.8 &4086.6 &3726.7    &3725.2($D^0D^+$)    &3734.4          \\
                              &$1(1^+)$   &3844.8 &4133.3 &3844.7    &3843.2($D^0D^{*+}$)    &3871.8            \\
                              &$1(2^+)$   &3962.8 &4159.4 &3962.7    &3961.2($D^{*0}D^{*+}$)    &4017.3            \\
                              \hline
\multirow{3}{*}{$T_{bb}$}       &$1(0^+)$   &10562.3 &10730.2 &10562.3    &10561.6($B^{-}\bar{B}^0$)    &10558.5    \\
                              &$1(1^+)$   &10601.2 &10741.1 &10601.1    &10600.4($B^{-}\bar{B^*}^0$)    &10604.5     \\
                              &$1(2^+)$   &10639.9 &10751.6 &10639.9    &10639.2($B^{*-}\bar{B^*}^0$)    &10650.5    \\
\hline\hline
\end{tabular}
\end{center}
\end{table}

\begin{table*}[!t]
\linespread{1.2}
\begin{center}
\caption{ \label{boundresults2} The mass of iso-scalar $T_{cc}$ states with $I(J^P)=0(1^+)$ (in unit of MeV). }\setlength{\tabcolsep}{8mm}{
\begin{tabular}{ccccc}
\hline\hline\noalign{\smallskip}
Channel                                           &color structure             &$E$      &$E_{Theo}$               &$\Delta E$  \\
\hline
$[(c\bar{u})^0_1(c\bar{d})^1_1]^1$                &$1\times1$                  &3843.8   &3843.2 ($D^0D^{*+}$)      &+0.6    \\
$[(c\bar{u})^0_8(c\bar{d})^1_8]^1$                &$8\times8$                  &4168.7   &                         &+325.5  \\
                                                  &$1\times1$ + $8\times8$    &3842.0   &                         &-1.2   \\
$[(c\bar{u})^1_1(c\bar{d})^0_1]^1$                &$1\times1$                  &3843.8   &3843.2 ($D^{*0}D^{+}$)    &+0.6    \\
$[(c\bar{u})^1_8(c\bar{d})^0_8]^1$                &$8\times8$                  &4168.7   &                         &+325.5  \\
                                                  &$1\times1$ + $8\times8$    &3842.0   &                         &-1.2     \\
$[(c\bar{u})^1_1(c\bar{d})^1_1]^1$                &$1\times1$                  &3961.9   &3961.2 ($D^{*0}D^{*+}$)   &+0.7    \\
$[(c\bar{u})^1_8(c\bar{d})^1_8]^1$                &$8\times8$                  &4102.3   &                         &+141.1  \\
                                                  &$1\times1$ + $8\times8$    &3958.7   &                         &-2.5    \\
                                                  & all $1\times 1$ mixing    &3841.7 &3843.2 ($D^0D^{*+}$)        &-1.5 \\
                                                  &MM structure mixing            &3841.4   &3843.2 ($D^0D^{*+}$)      &-1.8     \\
$[(cc)^0_6(\bar{u}\bar{d})^1_{\bar{6}}]^1$        &$6\times \bar{6}$           &4115.1   &3843.2 ($D^0D^{*+}$)      &+271.9  \\
$[(cc)^1_{\bar{3}}(\bar{u}\bar{d})^0_3]^1$        &$\bar{3}\times 3$           &3704.8   &                         &-138.4  \\
                                                  &DA structure mixing            &3700.9   &                         &-142.3  \\
                                                  &MM and DA  mixing              &3660.7   &3843.2 ($D^0D^{*+}$)      &-182.5  \\
\hline\hline
\end{tabular}}
\end{center}
\end{table*}

\begin{table*}[!t]
\linespread{1.2}
\begin{center}
\caption{ \label{boundresults3} The mass of iso-scalar $T_{bb}$ states with $I(J^P)=0(1^+)$ (in unit of MeV). }\setlength{\tabcolsep}{8mm}{
\begin{tabular}{ccccc}
\hline\hline\noalign{\smallskip}
Channel                                           &color structure             &$E$      &$E_{Theo}$               &$\Delta E$  \\
\hline
$[(b\bar{u})^0_1(b\bar{d})^1_1]^1$                &$1\times1$                  &10590.3   &10600.4 ($B^{-}\bar{B^*}^0$)      &-10.1     \\
$[(b\bar{u})^0_8(b\bar{d})^1_8]^1$                &$8\times8$                  &10765.7   &                         &+165.3  \\
                                                  &$1\times1$ + $8\times8$     &10566.7   &                         &-33.7   \\
$[(b\bar{u})^1_1(b\bar{d})^0_1]^1$                &$1\times1$                  &10590.3   &10600.4 ($B^{*-}\bar{B}^0$)    &-10.1    \\
$[(b\bar{u})^1_8(b\bar{d})^0_8]^1$                &$8\times8$                  &10765.7   &                         &+165.3  \\
                                                  &$1\times1$ + $8\times8$     &10566.7   &                         &-33.7     \\
$[(b\bar{u})^1_1(b\bar{d})^1_1]^1$                &$1\times1$                  &10629.8   &10639.2 ($B^{*-}\bar{B^*}^0$)   &-9.4    \\
$[(b\bar{u})^1_8(b\bar{d})^1_8]^1$                &$8\times8$                  &10738.5   &                         &+99.3  \\
                                                  &$1\times1$ + $8\times8$     &10598.8   &                         &-40.4     \\
                                                & all $1\times 1$ mixing       &10551.8 &10600.4 ($B^{-}\bar{B^*}^0$) &-48.6 \\
                                                  &MM structure mixing            &10545.9   &10600.4 ($B^{-}\bar{B^*}^0$)      &-54.5     \\
$[(bb)^0_6(\bar{u}\bar{d})^1_{\bar{6}}]^1$        &$6\times \bar{6}$           &10746.8   &10600.4 ($B^{-}\bar{B^*}^0$)      &+146.4 \\
$[(bb)^1_{\bar{3}}(\bar{u}\bar{d})^0_3]^1$        &$\bar{3}\times 3$           &10298.9   &                         &-301.5  \\
                                                  &DA structure mixing            &$10298.2$   &                         &-302.2  \\
                                                  &                            &$10576.8$   &                         &-23.6  \\
                                                  &MM and DA  mixing              &$10282.7^{1st}$   &10600.4 ($B^{-}\bar{B^*}^0$)      &-317.7  \\
                                                  &                            &$10516.7^{2nd}$   &                                  &-83.7  \\
\hline\hline
\end{tabular}}
\end{center}
\end{table*}

\subsection{Calculation of bound states}
The low-lying eigenvalues of iso-vector $T_{QQ}$ states with $I(J^P)=1(0^+), 1(1^+), 1(2^+)$ are shown in Table \ref{boundresults1}. From the table,
we can easily see that the low-lying energies are all higher than the corresponding theoretical thresholds no matter in meson-meson structure,
diquark-antidiquark structure or even considering the coupling of the two structures. No bound states are found.

Table \ref{boundresults2} and Table \ref{boundresults3} gives the mass of iso-scalar $T_{cc}$ and $T_{bb}$ with $I(J^P)=0(1^+)$, respectively.
The first column is the color-spin channel (list in Table \ref{bases}); the second column refers to the color structure including $1\times1$,
$8\times8$ and their mixing, as well as $\bar{3}\times3$, $6\times \bar{6}$ and their mixing. The following two columns represent the theoretical
mass ($E$) and theoretical thresholds ($E_{Theo}$). The last column gives the binding energy $\Delta E= E-E_{Theo}$. In Table \ref{boundresults2},
we found that: There exist no bound states in single channel calculation except for diquark-antidiquark $\bar{3}\times 3$ structure. When considering
the coupling of $1\times 1$ and $8 \times 8$ color structures in meson-meson configuration, a loosely bound state with mass $M=3841.4$ MeV is obtained
with the binding energy -1.8 MeV, which is consistent with the recently experiment data by LHCb. In diquark-antidiquark coupling calculations,
a tightly bound state with mass $M=3700.9$ MeV is get with the binding energy -142.3 MeV. We obtain the lowest state with mass $M=3660.7$ MeV,
a tightly bound state with binding energy -182.5 MeV, by considering the complete channels coupling effects both including meson-meson and
diquark-antidiquark structures.

The same calculations are extended to $T_{bb}$ system in Table \ref{boundresults3}. With the increasing of $m_{Q}$, the bound states are easier to get. When considering the single channel, for $1 \times 1$ and $\bar{3} \times 3$ color structures, there exist bound states, except for $8 \times 8$ and $6 \times \bar{6}$ color structures. The mixing of meson-meson structure leads to a bound state 10545.9 MeV, which has -54.5 MeV binding energy. The mixing of diquark-antidiquark structures obtains a deeper bound state. When considering the coupling of meson-meson and diquark-antidiquark structures, we found two bound states with mass 10282.7 and 10516.7 MeV.


Let us compare our results with some recent calculations in Table \ref{compare}.
From the table, we can see that for $T_{cc}$ of $0(1^+)$, controversy exists whether there are bound states. Our one shallow bound state with binding energy -1.8 MeV is consistent with the Refs. \cite{Yang:2019itm,Ren:2021dsi,Agaev:2021vur}, also in case of only meson-meson structures considered. In lattice QCD calculations, a few work support it is a bound state, also with small binding energy \cite{Junnarkar:2018twb}. If one considers diquark-antidiquark structures in the calculations, deeper bound state for $T_{cc}$ of $0(1^+)$ is also obtained such as \cite{Tan:2020ldi,Deng:2018kly}. For $T_{bb}$ of $0(1^+)$, the situation becomes much clear. Almost of the work obtained the bound results. In the case of considering the meson-meson structures, our result with binding energy -54.5 MeV is well consistent with the Refs. \cite{Yang:2019itm,Lu:2020rog} in the quark model and also the Ref. \cite{Bicudo:2012qt} in the lattice QCD calculation. No matter for $T_{cc}$ or $T_{bb}$, when diquark-antidiquark structures are taken into account, deeper binding energies will appear, which seems to be somewhat misleading. In some other work, same conclusions are obtained, such as \cite{Hyodo:2012pm,Li:2012ss,Richard:2018yrm,Meng:2020knc}.

In order to understand the mechanism of forming the bound states for different color structures and their couplings, the contributions of each term in the
Hamiltonian for $T_{cc}$ with $0(1^+)$ are given in Table \ref{contributions}. In the table, $K_1$ and $K_2$ is the kinetic energy of two sub-clusters,
and $K_3$ represents the relative kinetic energy between these two clusters. $V^{\rm{C, Coul, CMI}}$ is on behalf of the confinement (C), color coulomb (Coul), color-magnetic interaction (CMI). $V^{\eta, \pi, \sigma}$ represents pseudoscalar meson $\eta$, $\pi$ exchange and scalar meson $\sigma$ exchange.
To understand the mechanism, one has to compare the results in the tetraquark calculation to those of two free mesons which are listed in the last column in Table \ref{contributions}.
For color singlet channels (index 1,3,5), the attractions provided by $\pi$ and $\sigma$ are too weak to overcome the relative kinetic energy to
bind two color singlet clusters together, and these channels are scattering states when they stand alone. Coupling these color singlet channels adds
more attraction, a shallow bound state is formed (see the column headed by 1+3+5). The effect of coupling color octet-octet channels is very small, adding
$-0.3$ MeV binding energy (see the column headed by MM). If only the meson-meson structure is considered, the reported state by LHCb Collaboration can
be explained by the molecular state. However, for multi-quark system, more color structures are possible, in which the diquark-antidiquark structure
with color representations $\bar{3} \times 3$ and $6 \times \bar{6}$ is often invoked. For $\bar{3} \times 3$ channel (column headed by 8), the strong
attractions from $\pi$ meson exchange, $\sigma$ meson exchange and CMI overcome the repulsion from the kinetic energy and bind two
``good diquark/antidiquark" \cite{diquark} to form a deep bound state. The channel coupling between $\bar{3} \times 3$ and $6 \times \bar{6}$ channels
adds a little more binding energy to the system. The attraction from $\pi$-meson exchange in ``good diquark" is very large
because of its compact structure, so does the color magnetic interaction. The calculations without invoking the pseudoscalar exchange do not obtain bound state of $T_{cc}$ \cite{EPJA19} support this statement.

\begin{table*}[!t]
\setlength\tabcolsep{0.3pt}
\linespread{1.2}
\begin{center}
\caption{ \label{compare} The results of $T_{cc}$ and $T_{bb}$ tetraquarks with $0(1^+)$ in comparison with the other theoretical calculations (in unit of MeV).
N stands for "no bound state''. In the second column, (MM) represents only meson-meson structures considered. (DA) stands for only diquark-antidiquark configurations. (MM+DA) represents the coupling of meson-meson and diquark-antidiquark structures.}
\begin{tabular}{ccccccccccccccccccccc}
\hline\hline\noalign{\smallskip}
$T_{cc}$ &This work   &\cite{Lee:2009rt}   &\cite{Luo:2017eub}   &\cite{Guo:2021yws}   &\cite{Karliner:2017qjm} &\cite{wpark} &\cite{Yang:2019itm}  &\cite{Tan:2020ldi}  &\cite{Lu:2020rog}    &\cite{Ebert:2007rn} &\cite{Ren:2021dsi}   &\cite{Meng:2020knc}  &\cite{Zhang:2021yul} &\cite{Noh:2021lqs}  &\cite{Deng:2018kly} &\cite{Agaev:2021vur}    &\cite{Ikeda:2013vwa}    &\cite{Junnarkar:2018twb} \\
         &-1.8 (MM)    &-79 &-96  &-98 &N     &N     &-1   &-182.3 &N     &N  &-1  &-23     &N &N &-150 &-3    &N  &-23.3$\pm$11.4   \\
         &-142.3 (DA)   &&&&&&&&&&& \\
         &-182.5 (MM+DA) &&&&&&&&&&& \\
         \hline
$T_{bb}$ &This work &\cite{Guo:2021yws} &\cite{Karliner:2017qjm} &\cite{Hernandez:2019eox} &\cite{Yang:2019itm} &\cite{Yang:2009zzp} &\cite{Tan:2020ldi} &\cite{Lu:2020rog} &\cite{Ebert:2007rn} &\cite{Meng:2020knc} &\cite{Zhang:2021yul} &\cite{Noh:2021lqs} &\cite{Deng:2018kly} &\cite{Bicudo:2012qt} &\cite{Francis:2016hui} &\cite{Junnarkar:2018twb} &\cite{Leskovec:2019ioa} &\cite{Mohanta:2020eed}  \\
       &-54.5 (MM) &-268.6 &-215 &-150 &-35 &-120.9 &-317 &-54 &-102 &-173 &N &-145 &-278 &-30 $\sim$ -57 &-189$\pm$13 &-143.3$\pm$34 &-128$\pm$24$\pm$10 &-167$\pm$19 \\
       &-302.2 (DA) &&&&&&&&&&& \\
       &$-317.7^{1st}$ (MM+DA)&&&&&&&&&&& \\
       &$-83.7^{2nd}$ (MM+DA) &&&&&&&&&&& \\
\hline\hline
\end{tabular}
\end{center}
\end{table*}

From quantum mechanics, one knows that the physical states are the linear combinations of all possible channels in various structures. So the channel coupling with inclusion of different structures is needed. Clearly, the effect of this channel coupling will push down the lowest state further.
A deep bound state with binding energy $-182.5$ MeV is arrived. Unfortunately, the shallow bound state in meson-meson structure is pushed up the threshold and disappears. Experimentally, if there is only one weekly bound $T_{cc}$ state, diquark-antidiquark configurations should be abandoned, or the quark-quark interaction in diquark structure should be modified with the accumulation of experimental data.

In discussion of hadronic states, the electromagnetic interaction is almost neglected due to the small coupling
constant $\alpha \approx 1/137$.
However, for the weakly bound state, $E_{B}/M << 1$, the electromagnetic interaction may play a role. To investigate the role of electric Coulomb,
we add the Coulomb potential $\alpha\frac{q_iq_j}{r_{ij}}$ to the system Hamiltonian and solve the Schr\"{o}dinger equation in the same way as
describing above. The results show that the states are stable against the inclusion of the electric Coulomb interaction. The lowest eigen-energies are
changed to 3840.7 MeV, 3700.2 MeV and 3660.2 MeV in MM structure, DA structure and structure mixing.

In order to learn about the nature of doubly heavy tetraquark states, we calculate the distances between any quark and quark/antiquark, which are shown in Table \ref{rms}. Need to note that, for $T_{QQ}$, because the heavy quarks $Q$ are the identical particles, as well as the light antiquark $\bar{u}$ and $\bar{d}$,
the distance $r_{12}$ in the Table \ref{rms} actually is not the real value between the first quark $Q$ and the second antidiquark $\bar{u}$.
It should be the average:
\begin{eqnarray}
r_{12}^2&=&r_{34}^2=r_{14}^2=r_{23}^2  \nonumber  \\
&=&\frac{1}{4}(r_{Q^1\bar{u}^2}^2+r_{Q^1\bar{d}^4}^2+r_{Q^3\bar{u}^2}^2+ r_{Q^3\bar{d}^4}^2 ).
\end{eqnarray}
The numbers '1,2,3,4' has been shown in Fig. \ref{structure}. From the table, for the weakly bound state of $T_{cc}$ with mass $3841.4$ MeV, just meson-meson structure considered, the distance $r_{13}$ and $r_{24}$ between two sub-clusters is larger than distance within one cluster, which indicates it is a molecule $DD^*$ state. On the contrary, the tightly bound states 3700.9 MeV and 3660.7 MeV are the diquark-antidiquark structures. Besides, from the table we can see that although the $\bar{u}\bar{d}$ is a good diquark, its size is sightly larger than that of the $cc$ diquark. This should be from the color-electric interaction, not the color-magnetic interaction.

\begin{table*}[!t]
\linespread{1.2}
\begin{center}
\caption{ \label{contributions} Contributions of each potential in the system Hamiltonian of $T_{cc}$ with $I(J^P)=0(1^+)$ for different color structures and their couplings. In the first row, to save the space, we give the labels in sequence of 1 $\sim$ 8 for eight channels of $T_{cc}$ with $0(1^+)$ list in Table \ref{bases}(in unit of MeV). }
\begin{tabular}{cccccccccccccc}
\hline\hline\noalign{\smallskip}
Channel             &1      &2       &3      &4      &5     &6      &7      &8      &$1+3+5$     &MM      &DA       &MM+DA  &$D^0+D^{*+}$  \\
\hline
$K_1$               &516.8  &317.2   &359.8  &322.8  &359.6 &347.3  &119.4  &191.2  &442.2                       &440.2   &192.2    &321.8    &516.7        \\
$K_2$               &359.8  &322.8   &516.8  &317.2  &359.6 &347.3  &295.4  &1052.2 &442.2                       &440.2   &1050.6   &929.7    &359.3           \\
$K_3$               &3.4    &307.3   &3.4    &307.3  &3.0   &323.4  &395.4  &262.3  &17.4                       &25.5    &267.0    &189.9     &0            \\
$V^{\rm{C}}$        &-390.4 &-264.2  &-390.4 &-264.2 &-335.8&-300.8 &-271.2 &-412.9 &-393.1                       &-393.6  &-415.6   &-432.8  &-390.0      \\
$V^{\rm{Coul}}$     &-612.9 &-517.8  &-612.9 &-517.8 &-543.8&-539.1 &-506.4 &-621.6 &-614.8                       &-614.7  &-623.6   &-645.8  &-612.7      \\
$V^{\rm{CMI}}$      &-113.1 &-6.6    &-113.1 &-6.6   &38.7  &-77.9  &-6.8   &-350.6 &-118.7                       &-120.1  &-356.7   &-346.4  &-112.1      \\
$V^{\eta}$          &0      &5.5    &0       &5.5    &0      &6.5   &-2.5   &86.7   &0.9                           &1.1     &86.2    &75.4     &0           \\
$V^{\pi}$           &-0.5   &-56.9  &-0.5    &-56.9  &-0.4   &-63.9 &28.3   &-530.6 &-10.8                          &-12.9   &-527.4  &-464.6   &0           \\
$V^{\sigma}$        &-1.3   &-20.5  &-1.3    &-20.5  &-1.1   &-22.2 &-18.5  &-53.8  &-5.6                          &-6.1    &-53.8   &-48.8    &0           \\
$E$                 &3843.8 &4168.7  &3843.8 &4168.7 &3961.9&4102.3 &4115.1 &3704.8 &3841.7                       &3841.4  &3700.9  &3660.7   &3843.2       \\
$\Delta E$          &+0.6   &+325.5  &+0.6   &+325.5 &+0.7  &+141.1 &+271.9 &-138.4 &-1.5                       &-1.8    &-142.3   &-182.5    &0           \\
\hline\hline
\end{tabular} \end{center}

\begin{center}
\caption{ \label{rms} The distance between any quark and quark/antiquark of the bound states for $T_{QQ}$ with $0(1^+)$ in meson-meson structures (MM), diquark-antidiquark structures (DA) and in complete channel couplings (MM+DA). }
\begin{tabular}{ccccccccc}
\hline\hline\noalign{\smallskip}
          &Color structure &Bound state &$r_{12}$   &$r_{34}$   &$r_{13}$     &$r_{24}$   &$r_{14}$   &$r_{23}$  \\
\hline
\multirow{3}{*}{$T_{cc}$} & MM   &3841.4       &1.861      &1.861       &2.496       &2.607      &1.861      &1.861     \\
                          & DA   &3700.9       &0.684      &0.684       &0.517       &0.595      &0.684      &0.684     \\
                          & MM+DA &3660.7      &0.698      &0.698       &0.565       &0.670      &0.698      &0.698     \\
\hline
\multirow{4}{*}{$T_{bb}$} & MM   &10545.9     &0.671      &0.671       &0.458       &0.928      &0.671      &0.671     \\
                          & DA   &10298.2     &0.639      &0.639       &0.352       &0.591      &0.639      &0.639    \\
                          & MM+DA &$10282.7^{1st}$    &0.646      &0.646       &0.372       &0.616      &0.646      &0.646     \\
                          &      &$10516.7^{2nd}$   &0.756      &0.756       &0.722       &0.815      &0.756     &0.756     \\
\hline\hline
\end{tabular}
\end{center}
\end{table*}

\subsection{Calculation of resonance states}

\begin{figure}
\center{\includegraphics[width=7.0cm]{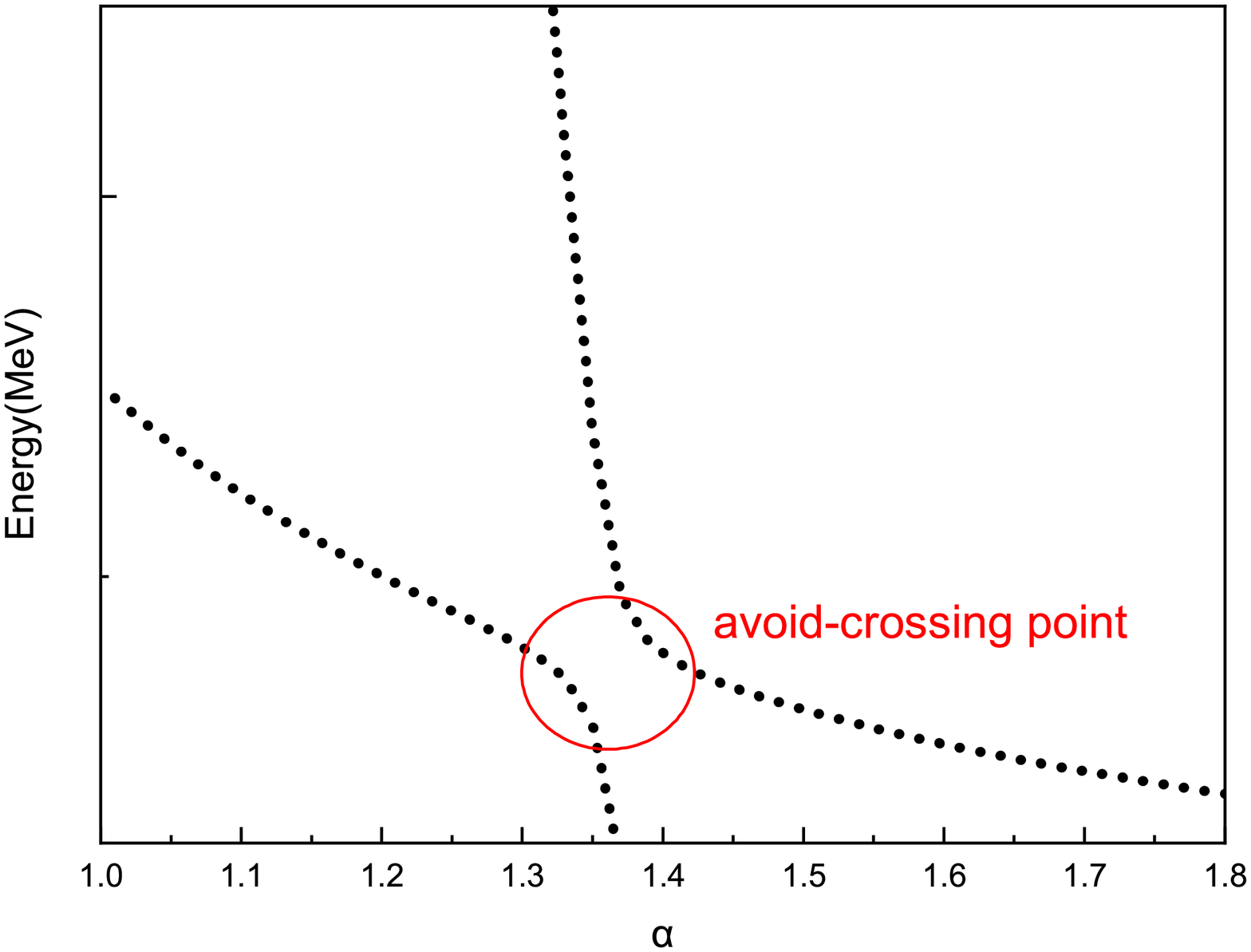}}
\caption{\label{avoid-crossing} Stabilization graph for the resonance.}
\end{figure}

\begin{figure}
\center{\includegraphics[width=9.0cm]{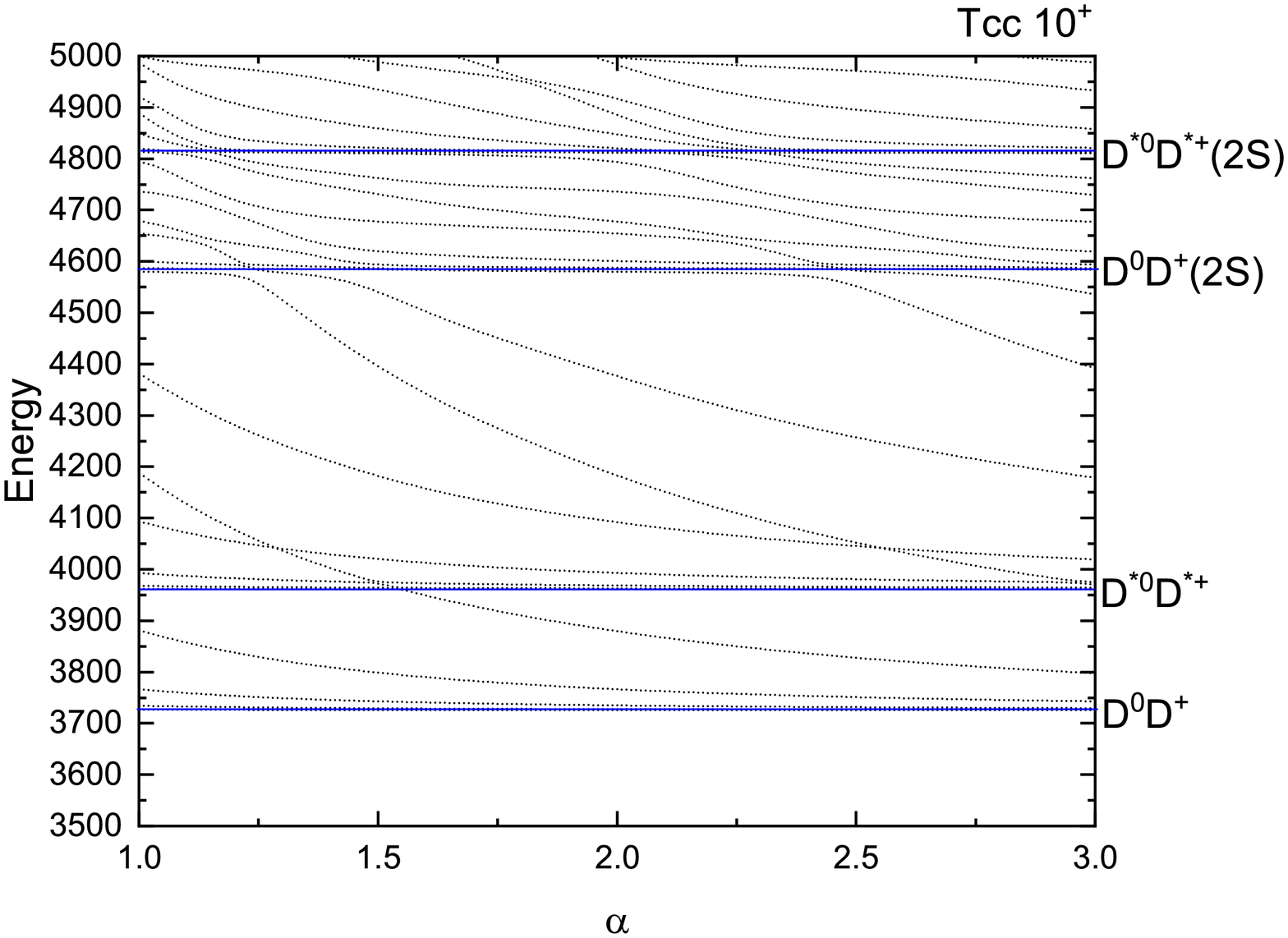}}
\caption{\label{tcc10} The stabilization plots of the energies of $T_{cc}$ states for $I(J^P)=1(0^+)$ with respect to the scaling factor $\alpha$.}
\end{figure}

\begin{figure}
\center{\includegraphics[width=9.0cm]{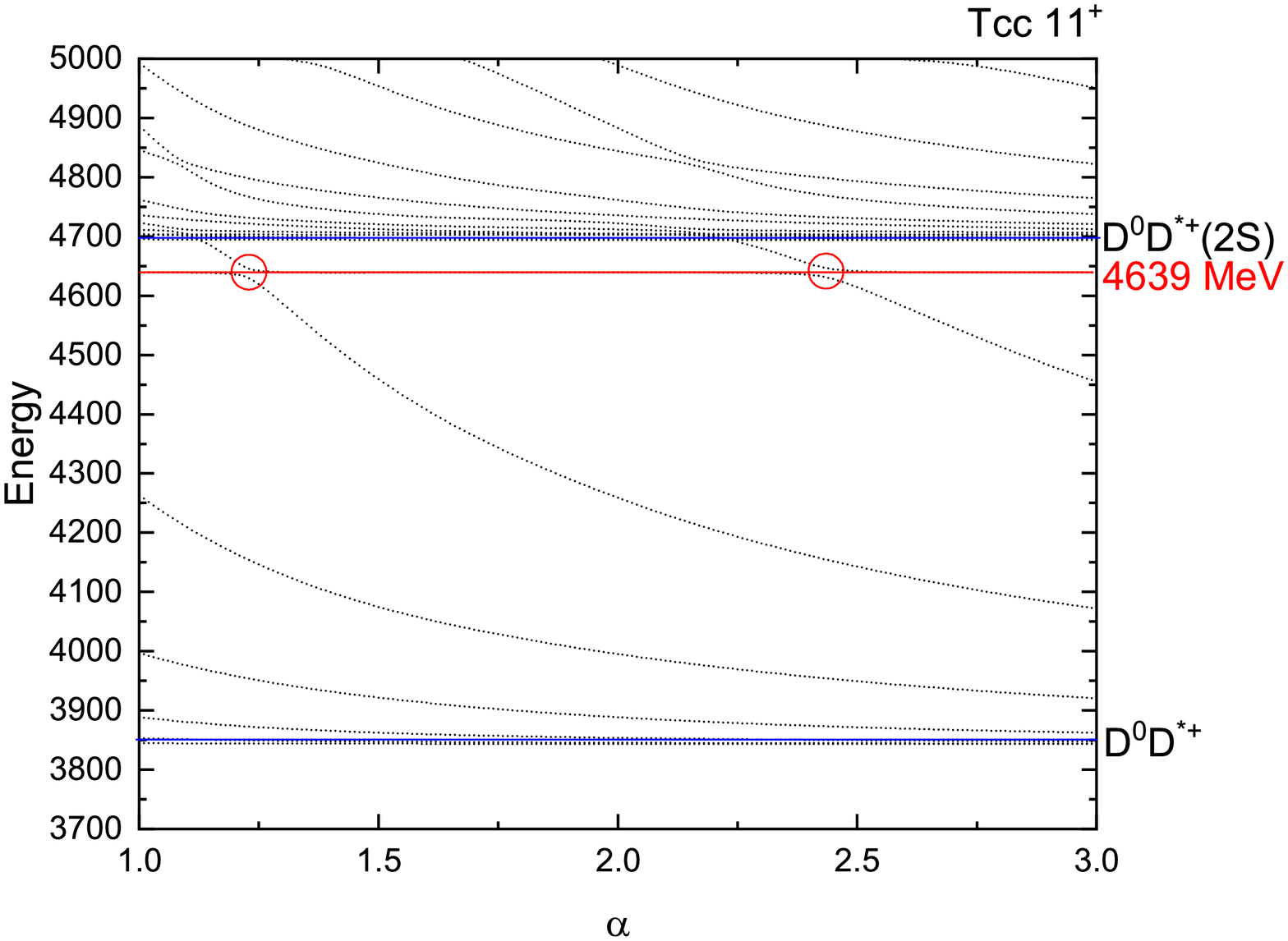}}
\caption{\label{tcc11} The stabilization plots of the energies of $T_{cc}$ states for $I(J^P)=1(1^+)$ with respect to the scaling factor $\alpha$.}
\end{figure}

\begin{figure}
\center{\includegraphics[width=9.0cm]{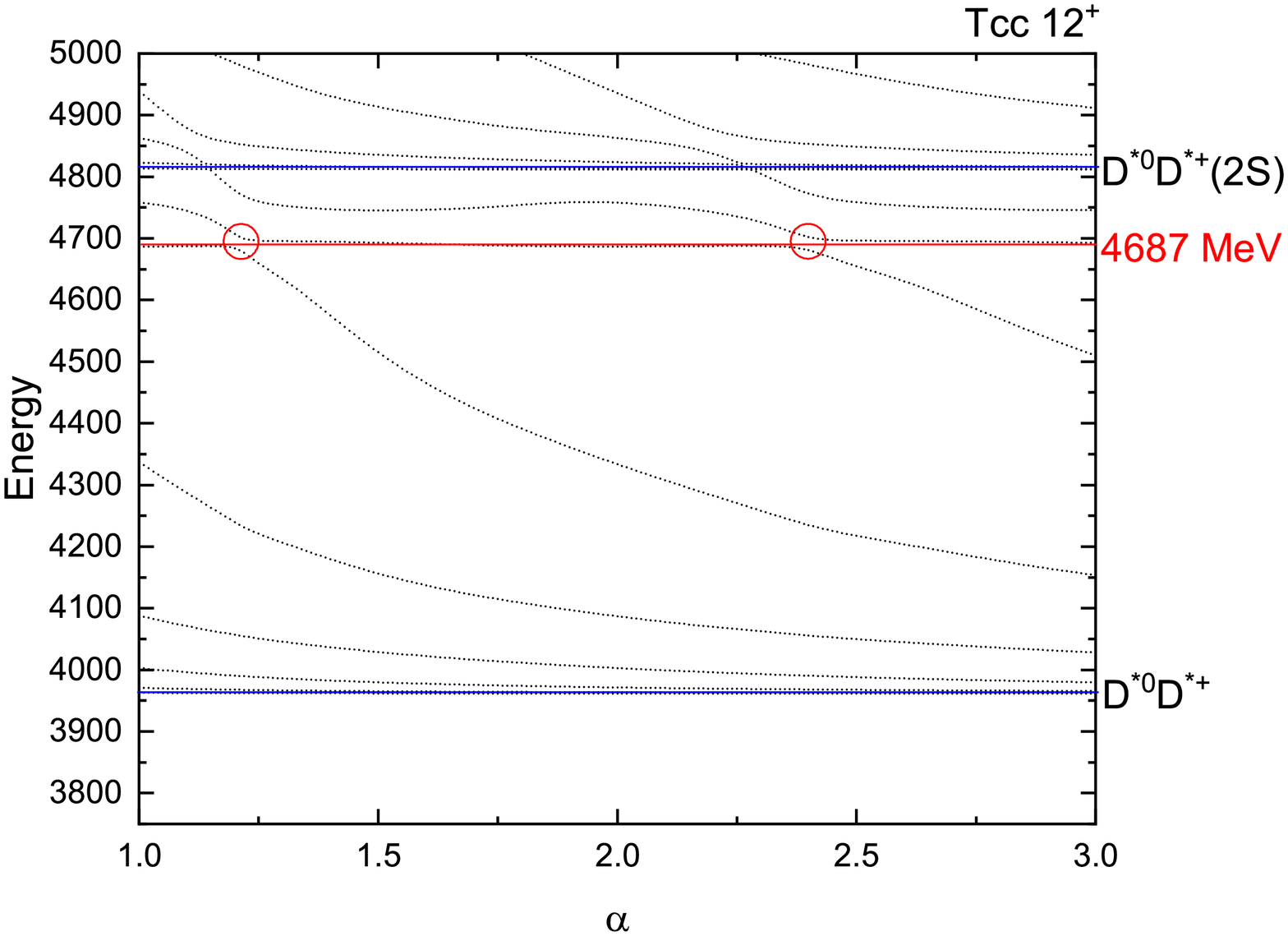}}
\caption{\label{tcc12} The stabilization plots of the energies of $T_{cc}$ states for $I(J^P)=1(2^+)$ with respect to the scaling factor $\alpha$.}
\end{figure}

\begin{figure}
\center{\includegraphics[width=9.0cm]{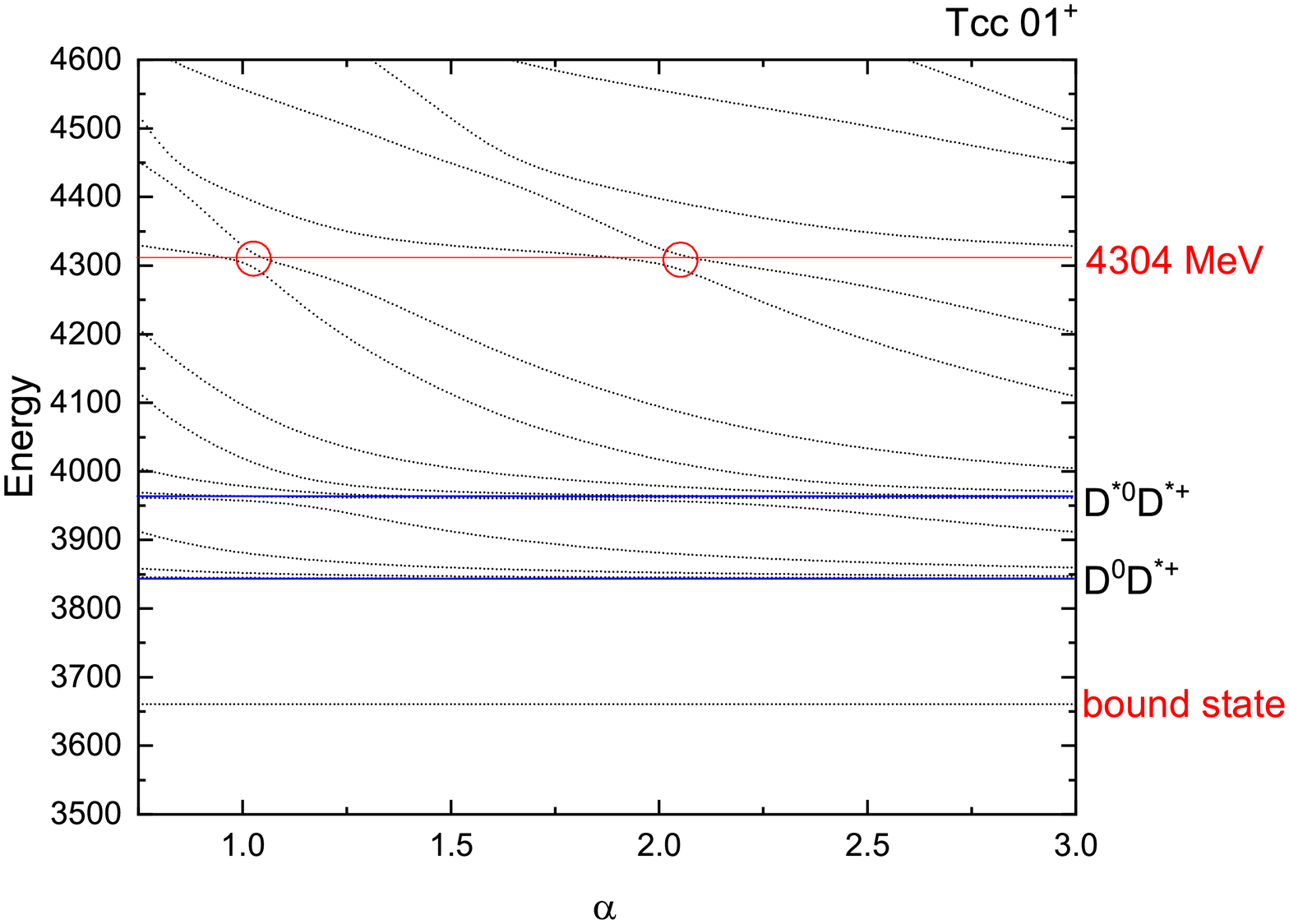}}
\caption{\label{tcc01} The stabilization plots of the energies of $T_{cc}$ states for $I(J^P)=0(1^+)$ with respect to the scaling factor $\alpha$.}
\end{figure}

\begin{figure}
\center{\includegraphics[width=9.0cm]{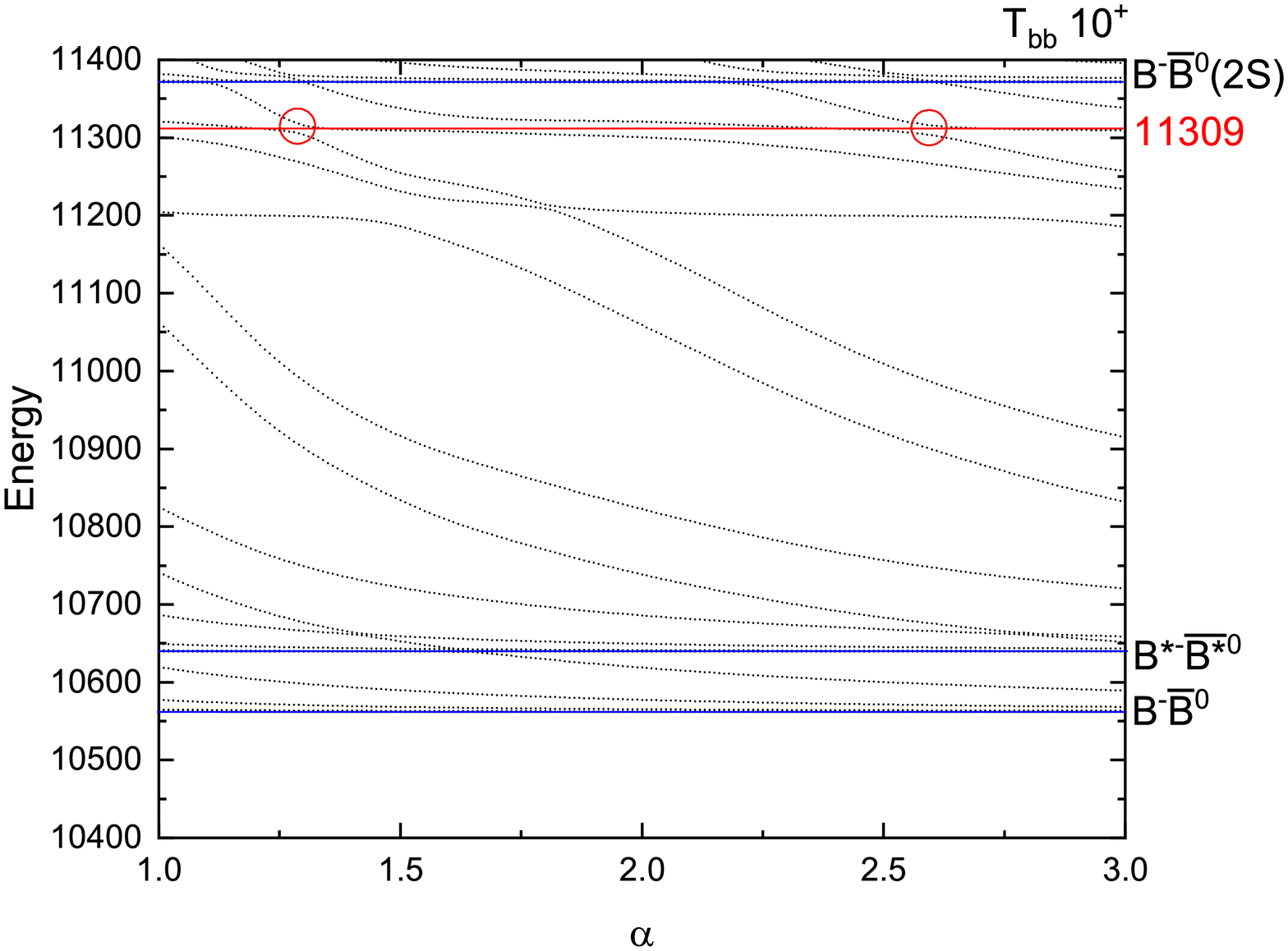}}
\caption{\label{tbb10} The stabilization plots of the energies of $T_{bb}$ states for $I(J^P)=1(0^+)$ with respect to the scaling factor $\alpha$.}
\end{figure}

\begin{figure}
\center{\includegraphics[width=9.0cm]{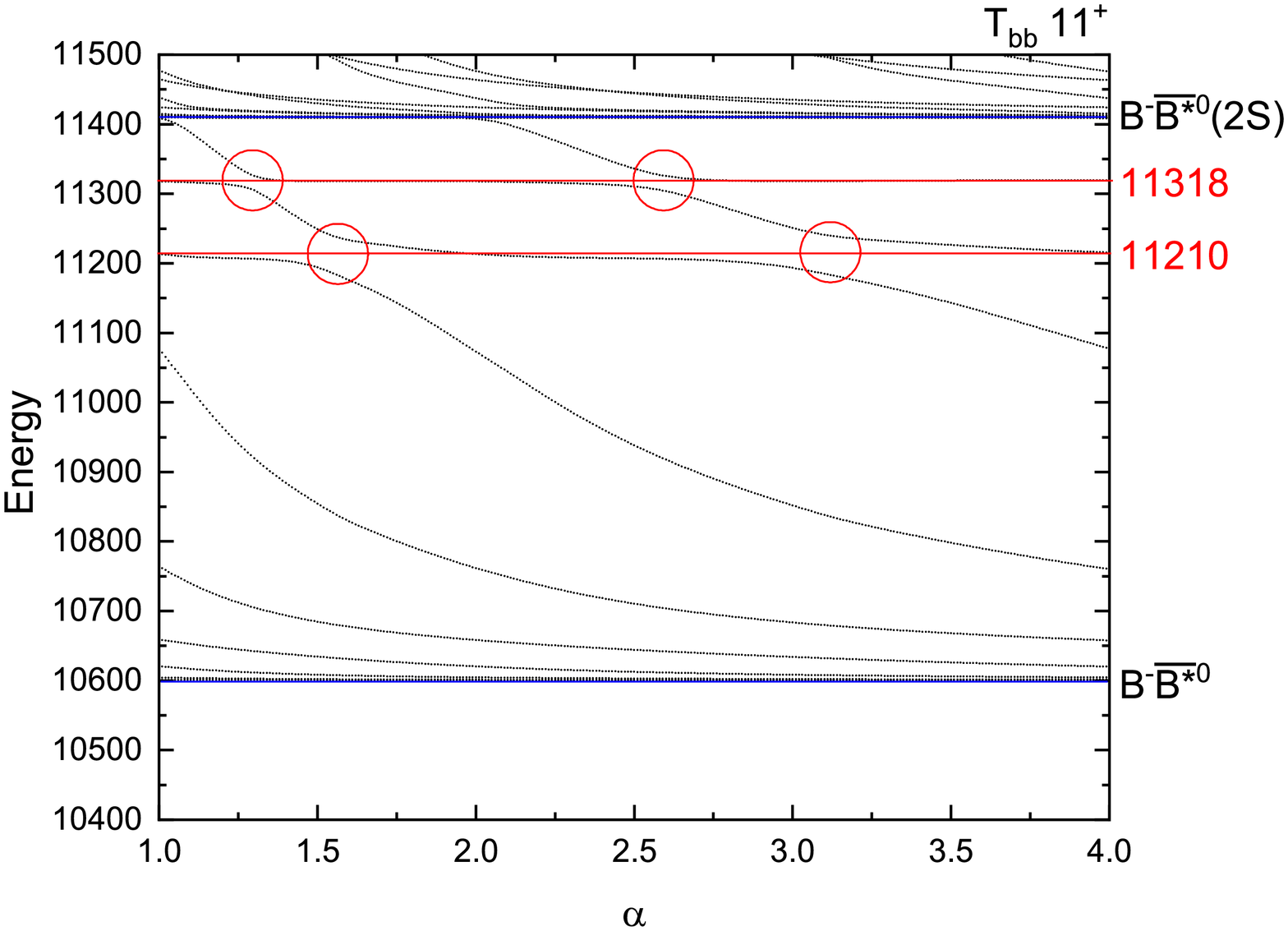}}
\caption{\label{tbb11} The stabilization plots of the energies of $T_{bb}$ states for $I(J^P)=1(1^+)$ with respect to the scaling factor $\alpha$.}
\end{figure}

\begin{figure}
\center{\includegraphics[width=9.0cm]{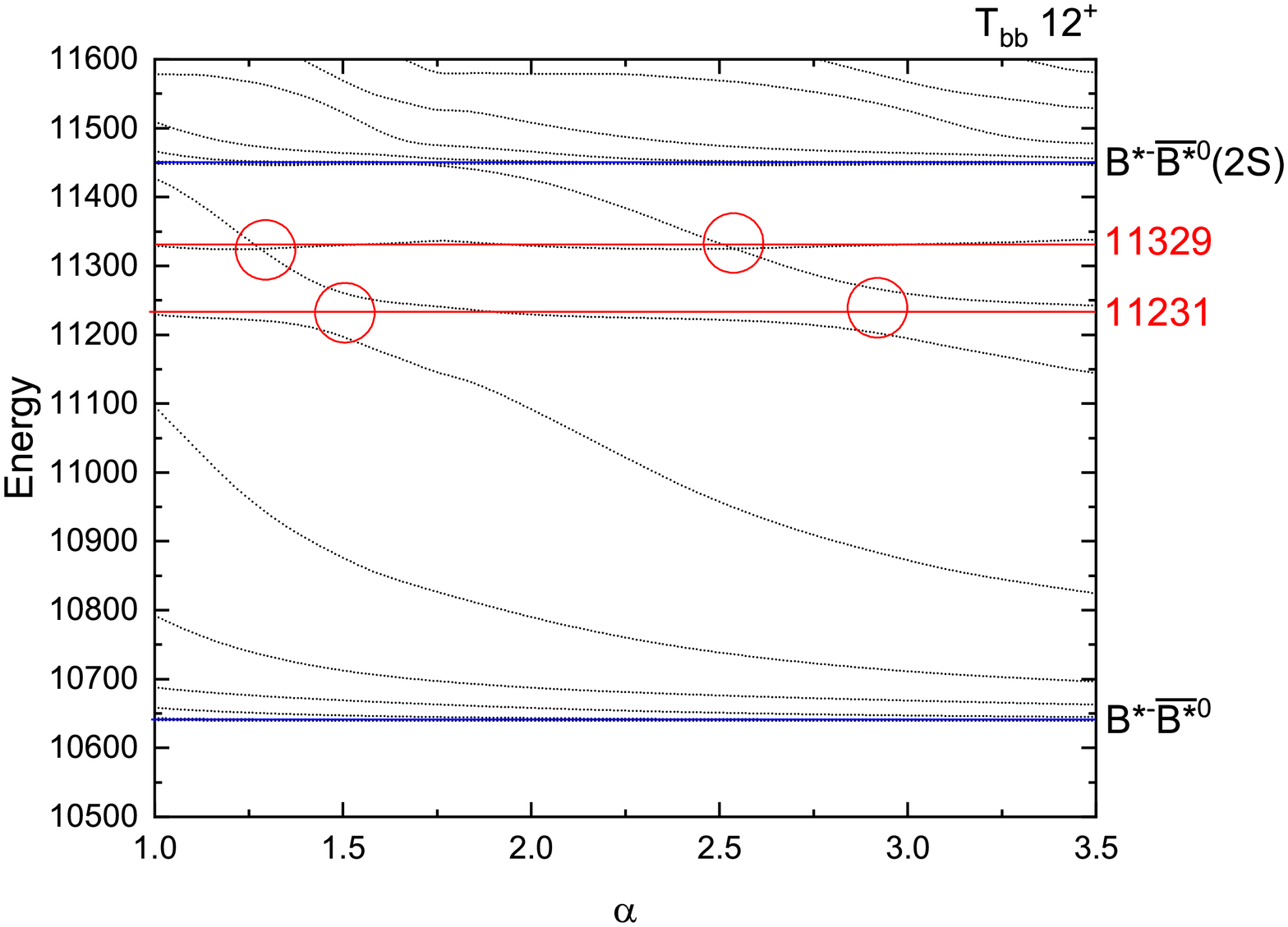}}
\caption{\label{tbb12} The stabilization plots of the energies of $T_{bb}$ states for $I(J^P)=1(2^+)$ with respect to the scaling factor $\alpha$.}
\end{figure}

\begin{figure}
\center{\includegraphics[width=9.0cm]{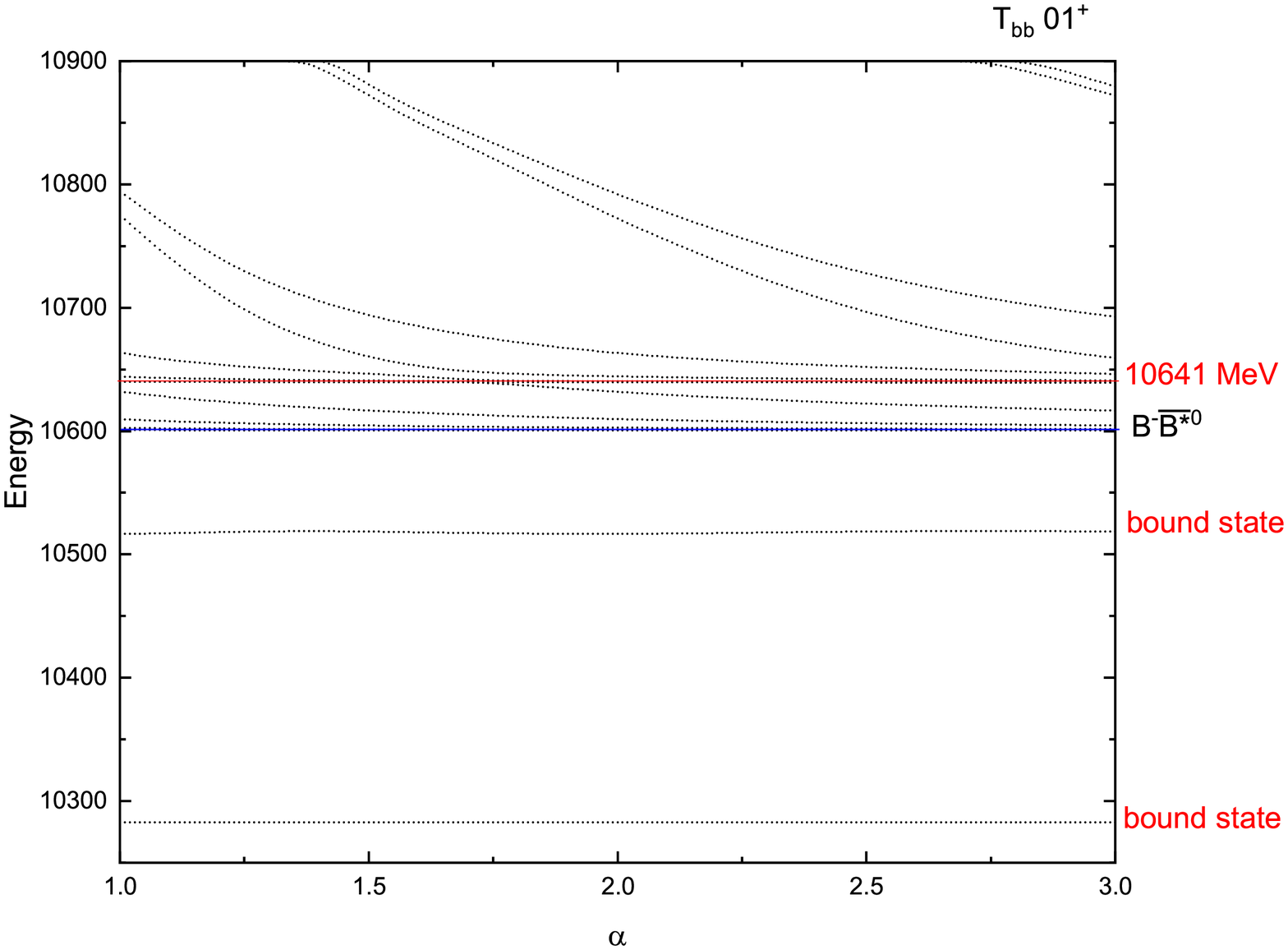}}
\caption{\label{tbb01} The stabilization plots of the energies of $T_{bb}$ states for $I(J^P)=0(1^+)$ with respect to the scaling factor $\alpha$.}
\end{figure}

In our work, we also concern about the possible resonance states of iso-vector and iso-scalar $T_{QQ}$. To find the genuine resonances, the dedicated real scaling (stabilization) method is employed. In our previous work \cite{Chen:2021uou}, this method is used successfully to explain the $Z_{cs}(3985)$ observed by BESIII Collaboration \cite{BESIII:2020qkh}. To realize the real scaling method in our calculation, we need multiply a factor $\alpha$ by the Gaussian size parameter $r_n$ in Eq. (\ref{gaussiansize}) just for the meson-meson structure with color singlet-singlet configuration. In our calculation, $\alpha$ takes the values from 1.0 to 3.0. With the increasing of $\alpha$, all sates will fall off towards its thresholds, but bound states should be stable and resonances will appear as an avoid-crossing structure just like Fig. \ref{avoid-crossing} \cite{rs1}. In the figure, at the avoid-crossing structure, there are two lines. The above one is a scattering state with larger slope, which will fall down to the threshold. The down line represents the resonance state, which tries to keep stable with smaller slope. When resonance state and scattering state interacts with each other, it brings about an avoid-crossing point in Fig. \ref{avoid-crossing}. With the increasing of the scaling factor $\alpha$, the energy of the other higher scattering state will fall down, interact with this resonance state again and another avoid-crossing point will emerge once more. In our work, the number of repetition of avoid-crossing points equals 2 owing to the large amount of computation. Theoretically speaking, with the continuously increasing of $\alpha$, the avoid-crossing point will appear in succession. Then we said we found a resonance state. For bound states, with the increasing of $\alpha$, they always stay stable. We show the results for $T_{QQ}$ system with all possible quantum numbers in Figs. (\ref{tcc10}-\ref{tbb01}) by considering the complete channel couplings.

Figure \ref{tcc10} represents the $T_{cc}$ states for $1(0^+)$. The four horizontal blue lines are the thresholds of $D^0D^+ (0 \otimes 0 \rightarrow 0)$, $D^{*0}D^{*+} (1 \otimes 1 \rightarrow 0)$ and their excited states $D^0D^+(2S)$ and $D^{*0}D^{*+}(2S)$. More higher energies are not list here. And we found no resonances for this state. From Fig. \ref{tcc11}, two thresholds $D^0D^{*+}$ and its excited states $D^0D^{*+}(2S)$ appear obviously. Around the energy of 4639 MeV, we found the avoid-crossing phenomenon, which is on behalf of a genuine resonance state. In Fig. \ref{tcc12}, we can see that the energy of the lowest resonance state is about 4687 MeV for $T_{cc}$ with $1(2^+)$. For iso-scalar $T_{cc}$ with $0(1^+)$, a bound state with mass 3660 MeV is obtained and the lowest resonance state shows with stable energy 4304 MeV in Fig. \ref{tcc01}.

For iso-vector and iso-scalar $T_{bb}$ states, there are also some resonance states found. They are states with the mass respectively 11309 MeV for $1(0^+)$, 11210 and 11318 MeV for $1(1^+)$, 11231 and 11329 MeV for $1(2^+)$, 10641 MeV for $0(1^+)$. There may be more resonance states with the higher energies, which may be too wide to be observed or too hard to be produced. So in our work, we only give the resonance states with as low as possible energies.

What's more, we calculated the decay widths of these resonance states using the formula taken from reference \cite{rs1},
\begin{align}
\Gamma=4|V(\alpha)|\frac{\sqrt{|S_r||S_c|}}{|S_c-S_r|},
\end{align}
where, $V(\alpha)$ is the difference between the two energies at the avoid-crossing point with the same value $\alpha$. $S_r$ and $S_c$
are the slopes of scattering line and resonance line, respectively. For each resonance, we get the values of the decay width at the first and the second
avoid-crossing point, and we finally give the average decay width of these two values. The results are shown in Table \ref{decaywidth}. It should be noted that the decay width is the partial strong two-body decay width to $S$-wave channels included in the calculation. For example, for $T_{cc}$ of $1(1^+)$, there is a resonance state with energy 4639 MeV, which has the decay width of 10 MeV in Table \ref{decaywidth}. This decay value is just the partial width to $D^0D^{*+}$ channel (see the threshold in Fig. \ref{tcc11}). For $T_{cc}$ of $0(1^+)$, the decay width of resonance(4304) with 38 MeV is the partial decay to $D^{*0}D^{*+}$, $D^0D^{*+}$ and $D^{+}D^{*0}$ channels (see the thresholds in Fig. \ref{tcc01}).

\begin{table}[!t]
\begin{center}
\caption{ \label{decaywidth} The decay widths of predicted resonances of $T_{cc}$ and $T_{bb}$ systems. (unit: MeV).}
\begin{tabular}{ccccc}
\hline\hline\noalign{\smallskip}
$I(J^P)$     &\multicolumn{2}{c}{$T_{cc}$}        &\multicolumn{2}{c}{$T_{bb}$}  \\
\hline
                   &Resonance states   &$\Gamma$        &Resonance states &$\Gamma$ \\
$1(0^+)$             &...               & ...               &11309            &23\\
$1(1^+)$             &4639               &10                &11210            &62\\
                        &&                              &11318            &21\\
$1(2^+)$             &4687               &9                &11231         &73\\
                        &&                              &11329         &0.4\\
$0(1^+)$              &4304               &38                &10641         &2\\
\hline\hline
\end{tabular}
\end{center}
\end{table}

\section{Summary}
\label{summary}
In the framework of the chiral quark model, we do a systematical calculation for the mass spectra for doubly heavy $T_{QQ}$ with quantum numbers $I(J^P)=1(0^+), 1(1^+), 1(2^+), 0(1^+)$ to look for possible bound states. Meanwhile, some resonance states are found with the real scaling method.

In bound state calculations, we analyze the effects of different color structures such as $1 \times 1$, $8 \times 8$ for meson-meson structures and $\bar{3} \times 3$, $6 \times \bar{6}$ for diquark-antidiquark structures, on the binding energy of $T_{QQ}$. The masses of states with isospin $I=1$ are all above the corresponding thresholds, leaving no space for bound states.
For $T_{cc}$ with $0(1^+)$, we found that in meson-meson structure, a loosely bound state 3841.1 MeV is obtained with $-1.8$ MeV binding energy, which is
consistent with the recent observed experiment by LHCb. But in diquark-antidiquark structures, CMI potential, $\pi$-exchange and $\sigma$ exchange offer more attractions and a tightly bound state with mass 3700 MeV is obtained. The couplings of meson-meson and diquark-antidiquark structures can not be neglected, which leads to more deeper bound states and destroys the loosely bound state.
For more heavy $T_{bb}$ system with $0(1^+)$, the same conclusions are obtained similar to the case of $T_{cc}$ and it looks easier to find more deeper bound states. For example, a bound state 10545.9 MeV with binding energy -54.5 MeV is obtained only meson-meson structures considered. If only diquark-antidiquark structures are taken into account, two bound states emerge with binding energy -302.2 MeV and -23.6 MeV.
When considering the coupling of meson-meson and diquark-antidiquark configurations, these two states will have more deeper binding energy with -317.7 MeV and -83.7 MeV.

For resonance state calculations, some resonances are found. For $T_{cc}$, the energies of the possible resonances are 4639 MeV for $1(1^+)$, 4687 MeV for $1(2^+)$, 4304 MeV for $0(1^+)$. For $T_{bb}$, the resonance energies are larger than $T_{cc}$, which takes 11309 MeV for $1(0^+)$, 11210 and 11318 MeV for $1(1^+)$, 11231 and 11329 MeV for $1(2^+)$, 10641 MeV for $0(1^+)$. Hopefully, our results in this work by the phenomenological framework of the chiral quark model need to be confirmed in the future high energy experiments.

\section{Acknowledgment}
This work is partly supported by the National Natural Science Foundation of China under Grants No. 11847145 and No. 11865019.


\begin{thebibliography}{10}
\bibitem{GEM}E. Hiyama, Y. Kino, M. Kamimura, Prog. Part. Nucl.
Phys. \textbf{51}, 223 (2003)

\bibitem{rs2}Emiko Hiyama, Atsushi Hosaka, Makoto Oka, Jean-Marc
Richard, Phys. Rev. C \textbf{98}, 045208 (2018)

\bibitem{LHCb:2021vvq}
R.~Aaij \textit{et al.} [LHCb],
[arXiv:2109.01038 [hep-ex]]

\bibitem{LHCb:2021auc}
R.~Aaij \textit{et al.} [LHCb],
[arXiv:2109.01056 [hep-ex]]

\bibitem{SELEX:2002wqn}
M.~Mattson \textit{et al.} [SELEX],
Phys. Rev. Lett. \textbf{89}, 112001 (2002)

\bibitem{SELEX:2004lln}
A.~Ocherashvili \textit{et al.} [SELEX],
Phys. Lett. B \textbf{628}, 18-24 (2005)

\bibitem{LHCb:2017iph}
R.~Aaij \textit{et al.} [LHCb],
Phys. Rev. Lett. \textbf{119}, no.11, 112001 (2017)

\bibitem{Ader:1981db}
J.~P.~Ader, J.~M.~Richard and P.~Taxil,
Phys. Rev. D \textbf{25}, 2370 (1982)

\bibitem{Heller:1986bt}
L.~Heller and J.~A.~Tjon,
Phys. Rev. D \textbf{35}, 969 (1987)

\bibitem{Carlson:1987hh}
J.~Carlson, L.~Heller and J.~A.~Tjon,
Phys. Rev. D \textbf{37}, 744 (1988)

\bibitem{Zouzou:1986qh}
S.~Zouzou, B.~Silvestre-Brac, C.~Gignoux and J.~M.~Richard,
Z. Phys. C \textbf{30}, 457 (1986)

\bibitem{lipkin}
H. J. Lipkin, Phys. Lett. B 172, 241 (1986)

\bibitem{Lee:2009rt}
S.~H.~Lee and S.~Yasui,
Eur. Phys. J. C \textbf{64}, 283-295 (2009)

\bibitem{Hyodo:2012pm}
T.~Hyodo, Y.~R.~Liu, M.~Oka, K.~Sudoh and S.~Yasui,
Phys. Lett. B \textbf{721}, 56-60 (2013)

\bibitem{Luo:2017eub}
S.~Q.~Luo, K.~Chen, X.~Liu, Y.~R.~Liu and S.~L.~Zhu,
Eur. Phys. J. C \textbf{77}, no.10, 709 (2017)

\bibitem{Li:2012ss}
N.~Li, Z.~F.~Sun, X.~Liu and S.~L.~Zhu,
Phys. Rev. D \textbf{88}, no.11, 114008 (2013)

\bibitem{Guo:2021yws}
T.~Guo, J.~Li, J.~Zhao and L.~He,
[arXiv:2108.10462 [hep-ph]]

\bibitem{Weng:2021hje}
X.~Z.~Weng, W.~Z.~Deng and S.~L.~Zhu,
[arXiv:2108.07242 [hep-ph]]

\bibitem{EPJA19}
J.~Vijande, F.~Fern\'andez, A.~Valcarce and B. Silvestre-Brac,
Eur. Phys. J. A \textbf{19}, 383 (2004)

\bibitem{Karliner:2017qjm}
M.~Karliner and J.~L.~Rosner,
Phys. Rev. Lett. \textbf{119}, no.20, 202001 (2017)

\bibitem{wpark}
W. Park, S. Noh and S. H. Lee,
Nucl. Phys. A 983, 1 (2019) [arXiv:1809.05257 [nucl-th]]

\bibitem{Hernandez:2019eox}
F.~Fern\'andez, J.~Vijande, A.~Valcarce and J.~M.~Richard,
Phys. Lett. B \textbf{800}, 135073 (2020)

\bibitem{Yang:2019itm}
G.~Yang, J.~Ping and J.~Segovia,
Phys. Rev. D \textbf{101}, no.1, 014001 (2020)

\bibitem{Yang:2009zzp}
Y.~Yang, C.~Deng, J.~Ping and T.~Goldman,
Phys. Rev. D \textbf{80}, 114023 (2009)

\bibitem{Tan:2020ldi}
Y.~Tan, W.~Lu and J.~Ping,
Eur. Phys. J. Plus \textbf{135}, no.9, 716 (2020)

\bibitem{Lu:2020rog}
Q.~F.~L\"u, D.~Y.~Chen and Y.~B.~Dong,
Phys. Rev. D \textbf{102}, no.3, 034012 (2020)

\bibitem{Ebert:2007rn}
D.~Ebert, R.~N.~Faustov, V.~O.~Galkin and W.~Lucha,
Phys. Rev. D \textbf{76}, 114015 (2007)

\bibitem{Ren:2021dsi}
H.~Ren, F.~Wu and R.~Zhu,
[arXiv:2109.02531 [hep-ph]]

\bibitem{Xing:2018bqt}
Y.~Xing and R.~Zhu,
Phys. Rev. D \textbf{98}, no.5, 053005 (2018)

\bibitem{Yan:2018gik}
X.~Yan, B.~Zhong and R.~Zhu,
Int. J. Mod. Phys. A \textbf{33}, no.16, 1850096 (2018)

\bibitem{Richard:2018yrm}
J.~M.~Richard, A.~Valcarce and J.~Vijande,
Phys. Rev. C \textbf{97}, no.3, 035211 (2018)

\bibitem{Carames:2011zz}
T.~F.~Carames, A.~Valcarce and J.~Vijande,
Phys. Lett. B \textbf{699}, 291-295 (2011)

\bibitem{Meng:2020knc}
Q.~Meng, E.~Hiyama, A.~Hosaka, M.~Oka, P.~Gubler, K.~U.~Can, T.~T.~Takahashi and H.~S.~Zong,
Phys. Lett. B \textbf{814}, 136095 (2021)

\bibitem{Wang:2021yld}
F.~L.~Wang and X.~Liu,
Phys. Rev. D \textbf{104}, no.9, 094030 (2021)
doi:10.1103/PhysRevD.104.094030
[arXiv:2108.09925 [hep-ph]]

\bibitem{Zhang:2021yul}
W.~X.~Zhang, H.~Xu and D.~Jia,
Phys. Rev. D \textbf{104}, no.11, 114011 (2021)
doi:10.1103/PhysRevD.104.114011
[arXiv:2109.07040 [hep-ph]]

\bibitem{Noh:2021lqs}
S.~Noh, W.~Park and S.~H.~Lee,
Phys. Rev. D \textbf{103}, 114009 (2021)
doi:10.1103/PhysRevD.103.114009
[arXiv:2102.09614 [hep-ph]]

\bibitem{Deng:2018kly}
C.~Deng, H.~Chen and J.~Ping,
Eur. Phys. J. A \textbf{56}, no.1, 9 (2020)
doi:10.1140/epja/s10050-019-00012-y
[arXiv:1811.06462 [hep-ph]]

\bibitem{Du:2012wp}
M.~L.~Du, W.~Chen, X.~L.~Chen and S.~L.~Zhu,
Phys. Rev. D \textbf{87}, no.1, 014003 (2013)

\bibitem{Wang:2017dtg}
Z.~G.~Wang and Z.~H.~Yan,
Eur. Phys. J. C \textbf{78}, no.1, 19 (2018)

\bibitem{Tang:2019nwv}
L.~Tang, B.~D.~Wan, K.~Maltman and C.~F.~Qiao,
Phys. Rev. D \textbf{101}, no.9, 094032 (2020)

\bibitem{Navarra:2007yw}
F.~S.~Navarra, M.~Nielsen and S.~H.~Lee,
Phys. Lett. B \textbf{649}, 166-172 (2007)

\bibitem{Agaev:2021vur}
S.~S.~Agaev, K.~Azizi and H.~Sundu,
[arXiv:2108.00188 [hep-ph]]

\bibitem{Agaev:2018khe}
S.~S.~Agaev, K.~Azizi, B.~Barsbay and H.~Sundu,
Phys. Rev. D \textbf{99}, no.3, 033002 (2019)

\bibitem{Agaev:2020mqq}
S.~S.~Agaev, K.~Azizi, B.~Barsbay and H.~Sundu,
Eur. Phys. J. A \textbf{57}, no.3, 106 (2021)

\bibitem{Agaev:2019qqn}
S.~S.~Agaev, K.~Azizi and H.~Sundu,
Phys. Rev. D \textbf{99}, no.11, 114016 (2019)

\bibitem{Agaev:2020dba}
S.~S.~Agaev, K.~Azizi, B.~Barsbay and H.~Sundu,
doi:10.1140/epja/s10050-020-00187-9
[arXiv:2001.01446 [hep-ph]]

\bibitem{Azizi:2021aib}
K.~Azizi and U.~\"Ozdem,
[arXiv:2109.02390 [hep-ph]]

\bibitem{Xin:2021wcr}
Q.~Xin and Z.~G.~Wang,
[arXiv:2108.12597 [hep-ph]]

\bibitem{Brown:2012tm}
Z.~S.~Brown and K.~Orginos,
Phys. Rev. D \textbf{86}, 114506 (2012)

\bibitem{Ikeda:2013vwa}
Y.~Ikeda, B.~Charron, S.~Aoki, T.~Doi, T.~Hatsuda, T.~Inoue, N.~Ishii, K.~Murano, H.~Nemura and K.~Sasaki,
Phys. Lett. B \textbf{729}, 85-90 (2014)

\bibitem{Bicudo:2012qt}
P.~Bicudo \textit{et al.} [European Twisted Mass],
Phys. Rev. D \textbf{87}, no.11, 114511 (2013)
doi:10.1103/PhysRevD.87.114511
[arXiv:1209.6274 [hep-ph]]

\bibitem{Bicudo:2015vta}
P.~Bicudo, K.~Cichy, A.~Peters, B.~Wagenbach and M.~Wagner,
Phys. Rev. D \textbf{92}, no.1, 014507 (2015)

\bibitem{Francis:2016hui}
A.~Francis, R.~J.~Hudspith, R.~Lewis and K.~Maltman,
Phys. Rev. Lett. \textbf{118}, no.14, 142001 (2017)

\bibitem{Junnarkar:2018twb}
P.~Junnarkar, N.~Mathur and M.~Padmanath,
Phys. Rev. D \textbf{99}, no.3, 034507 (2019)

\bibitem{Leskovec:2019ioa}
L.~Leskovec, S.~Meinel, M.~Pflaumer and M.~Wagner,
Phys. Rev. D \textbf{100}, no.1, 014503 (2019)

\bibitem{Hudspith:2020tdf}
R.~J.~Hudspith, B.~Colquhoun, A.~Francis, R.~Lewis and K.~Maltman,
Phys. Rev. D \textbf{102}, 114506 (2020)

\bibitem{Mohanta:2020eed}
P.~Mohanta and S.~Basak,
Phys. Rev. D \textbf{102}, no.9, 094516 (2020)
doi:10.1103/PhysRevD.102.094516
[arXiv:2008.11146 [hep-lat]]

\bibitem{Meng:2021jnw}
L.~Meng, G.~J.~Wang, B.~Wang and S.~L.~Zhu,
Phys. Rev. D \textbf{104}, no.5, 051502 (2021)
doi:10.1103/PhysRevD.104.L051502
[arXiv:2107.14784 [hep-ph]]

\bibitem{Ling:2021bir}
X.~Z.~Ling, M.~Z.~Liu, L.~S.~Geng, E.~Wang and J.~J.~Xie,
[arXiv:2108.00947 [hep-ph]]

\bibitem{Dong:2021bvy}
X.~K.~Dong, F.~K.~Guo and B.~S.~Zou,
Commun. Theor. Phys. \textbf{73}, no.12, 125201 (2021)
doi:10.1088/1572-9494/ac27a2
[arXiv:2108.02673 [hep-ph]]

\bibitem{Qin:2020zlg}
Q.~Qin, Y.~F.~Shen and F.~S.~Yu,
Chin. Phys. C \textbf{45}, 103106 (2021)

\bibitem{Ali:2018ifm}
A.~Ali, A.~Y.~Parkhomenko, Q.~Qin and W.~Wang,
Phys. Lett. B \textbf{782}, 412-420 (2018)


\bibitem{Ali:2018xfq}
A.~Ali, Q.~Qin and W.~Wang,
Phys. Lett. B \textbf{785}, 605-609 (2018)

\bibitem{Cheng:2020wxa}
J.~B.~Cheng, S.~Y.~Li, Y.~R.~Liu, Z.~G.~Si and T.~Yao,
Chin. Phys. C \textbf{45}, no.4, 043102 (2021)

\bibitem{Casimir} G. S. Bali, Phys. Rev. D \textbf{62}, 114503 (2000).

\bibitem{anticonfine} V. Dmitrasinovic, Phys. Rev. D \textbf{67}, 114007 (2003).

\bibitem{Chen:2016npt}
X.~Chen and J.~Ping,
Eur. Phys. J. C \textbf{76}, no.6, 351 (2016)

\bibitem{Chen:2018hts}
X.~Chen and J.~Ping,
Phys. Rev. D \textbf{98}, no.5, 054022 (2018)
doi:10.1103/PhysRevD.98.054022

\bibitem{Chen:2019vrj}
X.~Chen,
Phys. Rev. D \textbf{100}, no.9, 094009 (2019)

\bibitem{Valcarce:2005em}
A.~Valcarce, H.~Garcilazo, F.~Fern{\'a}ndez and P.~Gonzalez,
\newblock Rept. Prog. Phys. {\bf 68}, 965 (2005)

\bibitem{PDG}
P.A. Zyla \emph{et al.} (Particle Data Group), Prog. Theor. Exp. Phys. 083C01 (2020)

\bibitem{diquark} R. Jaffe, Phys. Rep. \textbf{409}, 1 (2005)

\bibitem{Chen:2021uou}
X.~Chen, Y.~Tan and Y.~Chen,
Phys. Rev. D \textbf{104}, no.1, 014017 (2021)

\bibitem{BESIII:2020qkh}
M.~Ablikim \textit{et al.} [BESIII],
Phys. Rev. Lett. \textbf{126}, no.10, 102001 (2021)

\bibitem{rs1}J. Simon, J. Chem. Phys. \textbf{75}, 2465 (1981)
\end{thebibliography}
\end{document}